\documentclass{ajceam-class}
\usepackage[english]{babel}
\title{Building a quantum-ready ecosystem}

\author[1,4]{Abhishek Purohit}
\author[1]{Maninder Kaur}
\author[2]{Zeki Can Seskir}
\author[3]{Matthew T. Posner}
\author[1]{Araceli Venegas-Gomez}

\affil[1]{QURECA (Quantum Resources and Careers), Glasgow, Scotland, United Kingdom}
\affil[2]{Institute for Technology Assessment and Systems Analysis, KIT, Karlstraße 11, Karlsruhe, Germany}
\affil[3]{Optonique, Québec, Canada}
\affil[4]{James Watt School of Engineering, Electronics and Nanoscale Engineering Division, University of Glasgow, Glasgow, United Kingdom}

\abstract{
The emergence of quantum technologies has led to groundbreaking advancements in computing, sensing, secure communications, and simulation of advanced materials with practical applications in every industry sector. The rapid advancement of the quantum technologies ecosystem has made it imperative to assess the maturity of these technologies and their imminent acceleration towards commercial viability. The current status of quantum technologies is presented and the need for a quantum-ready ecosystem is emphasised. Standard Quantum Technology Readiness Levels (QTRLs) are formulated and innovative models and tools are defined to evaluate the readiness of specific quantum technology. In addition to QTRLs, Quantum Commercial Readiness Levels (QCRLs) is introduced to provide a robust framework for evaluating the commercial viability and market readiness of quantum technologies. Furthermore, relevant indicators concerning key stakeholders, including government, industry, and academia are discussed and ethics and protocols implications are described, to deepen our understanding of the readiness for quantum technology and support the development of a robust and effective quantum ecosystem. 
}


\begin{document}

\maketitle
\thispagestyle{fancy}

\section{Introduction}
\firstword{Q}{uantum}
physics undoubtedly is one of the most successful scientific theories ever established in terms of the accuracy of its predictions. A global effort to use quantum phenomena like entanglement, superposition, and coherence to create radically new technologies has resulted in the verge of the second quantum revolution. We can already see quantum technology-based products prevailing in the market and an increasing number of companies \cite{Seskir2022}, organisations, governments \cite{qureca2023}, and individuals making an effort to build a global quantum ecosystem. The prevalent question now is how to use quantum technology and not when quantum technology makes its appearance. Organisations have often used the term "quantum-ready" to assess their current situation with regard to their ability to integrate quantum technology into their existing framework. But the applicability of the term has a wide range of implications in different case scenarios. Therefore, it is crucial to understand and define the term quantum-ready \cite{ey}. In this paper, we explore and provide a systematic understanding of quantum readiness, how to evaluate it, and how entities, organisations or individuals can become quantum-ready.
\\
To begin, we define what quantum technology is as a category of new technology in Section 2. After classifying emerging technologies, we delve into the term quantum readiness and the need to address it in Section 3. Then we establish the importance of being quantum-ready by applying the term to various innovation models in Section 4. This helps us in providing tools to understand quantum readiness and identify various indicators to take into consideration while evaluating quantum readiness. Finally, in Section 5 we explore the relevance and meaning of being quantum-ready for the key stakeholders in the ecosystem.

\section{Classifying emerging technologies}
To define quantum readiness, it is essential to understand the nature of quantum technologies. This can help to estimate the potential impact and implications of these technologies, as well as identify areas where further research and development are needed.

Quantum technologies find themselves classified as emerging technologies due to their current status of technological maturity and the nascent status of their practical applications. While their underlying principles, founded in quantum mechanics, have been established for over a century, the journey of translating these principles into practical, scalable, and reliable technologies is still in its early stages. Quantum computing, for example, is a field where significant research is ongoing to overcome key challenges, such as achieving fault tolerance \cite{bell}, error correction \cite{chia} and scalability \cite{seb}. Quantum communication and quantum sensing, while more advanced than quantum computing, still need to clear hurdles related to miniaturisation \cite{antonio, anton}, reliability, and integration into existing infrastructures. Moreover, the commercial viability of quantum technologies is yet to be fully realised. While there is clear potential and growing interest from industry, widespread, large-scale commercial applications of these technologies are still on the horizon.
One way to classify emerging technologies is based on their level of maturity \cite{rotolo,freeman}. Firstly, early emerging technologies are technologies that are still in the early stages of development and may not yet have a clear application or commercial potential. Secondly, emerging-growth technologies have moved beyond the early stages of development and have begun to show commercial potential. These may still be facing significant technical or regulatory challenges. Finally, growth technologies are the ones that have achieved significant commercial success and are likely to have a significant impact on their respective industries.
\begin{figure}[h]
    \centering
    \includegraphics[scale=0.086]{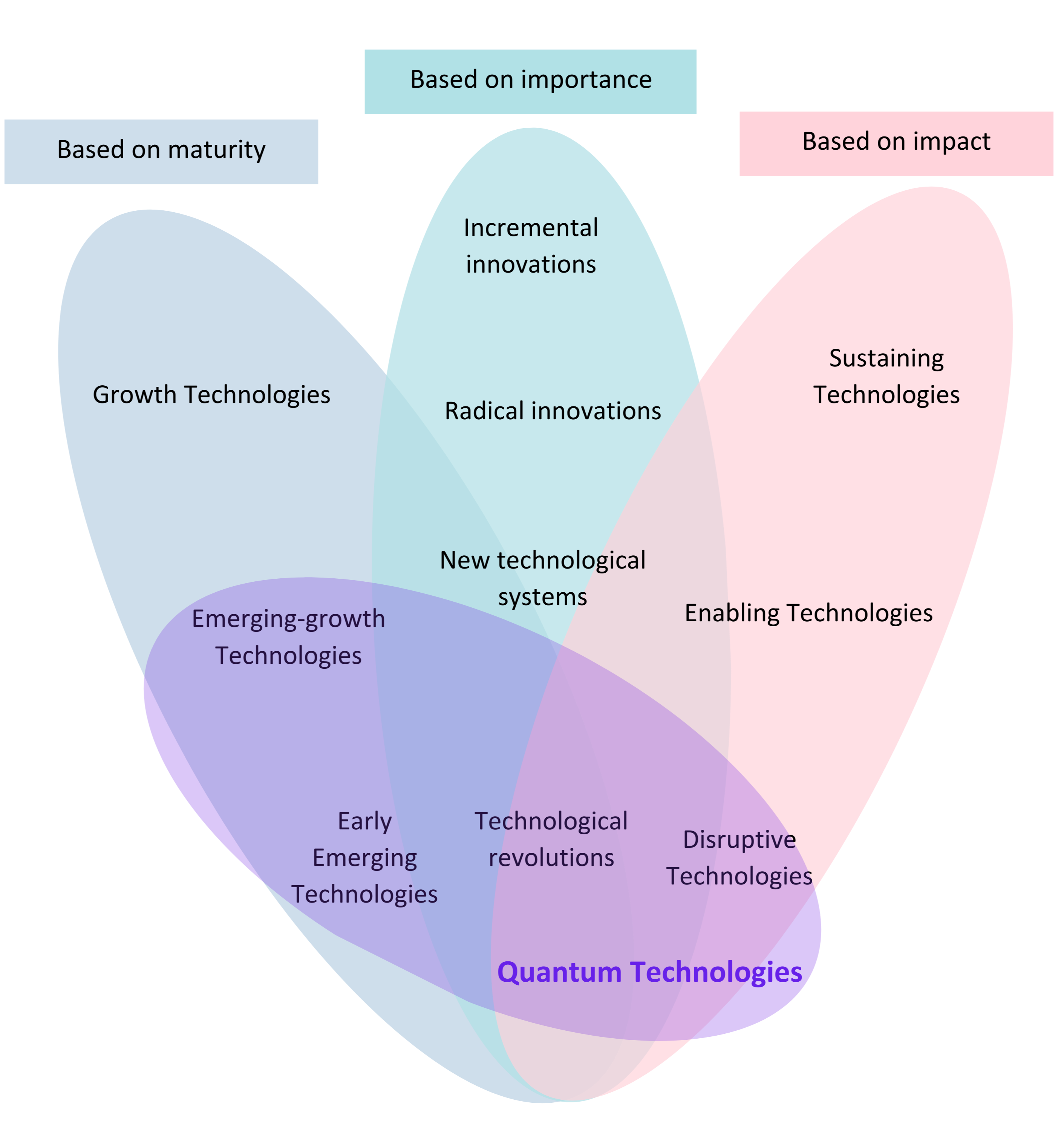}
    \vspace{0.2cm}
    \caption{Classification of emerging technologies and status of quantum technologies}
     
\end{figure}

Another way to classify emerging technologies would be based on their potential impact. 
Disruptive technologies are technologies that have the potential to fundamentally change the way an industry operates. They can create new markets or disrupt existing ones. When the term disruption is used, it generally indicates that an entirely novel method of operation has a substantial impact on how markets and industries currently perform \cite{miles}. Technology that supports the development of new goods, services, or business models without necessarily disrupting existing ones is known as enabling technology. Sustaining technologies improve the performance of existing products or services but do not create new markets or disrupt existing ones. 
Another taxonomy of innovations was proposed by Freeman that groups them according to increasing importance \cite{man}. This typology offers a conceptual framework for comprehending the contribution of quantum technology to a technological revolution. The typology consists of four types of innovations: incremental, radical, new technological systems, and technological revolutions. Smaller updates or changes made to the existing technology are referred to as incremental innovations. They may enhance the efficiency or quality of a product or service but do not fundamentally change its nature. Radical innovations involve important discoveries or developments that alter how a technology functions or is used. They frequently destroy established markets while offering new possibilities for expansion. The creation of completely new technological frameworks that integrate numerous related technologies or processes is called  new technology systems. These systems frequently result in the development of brand-new industries or the evolution of already-existing ones. Fundamental core discoveries resulting in the formation of a cluster of new technologies that may revolutionise a wide range of sectors constitute the technological revolutions. They are characterised by far-reaching and transformative impacts on various aspects of human life. 

Disruptive technologies, as coined by Clayton Christensen \cite{oro}, represent a transformative shift, a change in paradigms that disturb the status quo of existing markets or societal operations \cite{kumars}. This characteristic makes quantum technologies, with their potential to redefine numerous fields, a clear candidate for disruptive technologies \cite{coccia}. To comprehensively explore the potential of quantum technologies as disruptors, we can dissect them into three primary facets, each with historical parallels that underline their disruptive nature.
Firstly, we turn our attention to quantum computing. This cutting-edge computing paradigm, supported by quantum mechanics, bears a resemblance to the disruption caused by the invention of digital computers. The latter reshaped industries with an unprecedented surge in data processing speed and efficiency. Quantum computers, similarly, are poised to shatter current computational limits by accomplishing tasks at an exponential pace compared to their classical counterparts. The computational complexities within the scope of quantum computing, such as cryptographic problems, optimising logistical operations, and discovering new pharmaceutical compounds, to highlight a few, remain infeasible for traditional computing due to the exorbitant time-frames demanded for computation. Perhaps most disruptive is the impact on data security where in comparison to classical computers, quantum computers are capable of deciphering encryption codes in a fraction of the time.
Our second focal point is quantum communication technology, most notably Quantum Key Distribution (QKD). Poised to offer communication channels impervious to breaches, quantum communication has the potential to drastically transform the cybersecurity landscape. A close historical analogue is the proliferation of the internet, an innovation that fundamentally altered our modes of communication and information exchange. Finally, quantum sensing technologies, whose promise of ultra-precise measurements forecasts disruption across a multitude of sectors. GPS systems, medical technology, geological exploration, and more could see sweeping changes spurred by the sensitivity of quantum sensors. This potential disruption finds a historical echo in the introduction of radar technology, which indelibly imprinted its influence on navigation, meteorology, and warfare. The common thread across these potential disruptions is the departure from classical mechanics to quantum mechanics. This represents a paradigm shift as profound as the shift from mechanical to digital systems in the 20th century, which was a transition that disrupted and reshaped our civilisation. While the field of quantum technologies as a whole is classified as an emerging technology, certain subfields are progressing at a faster pace, nudging them into the realm of emerging-growth technologies. These sectors have advanced from the initial stages of research and development to demonstrate tangible commercial potential putting them in the emerging-growth technologies category (Figure 1). By transcending the limits of classical technologies and enabling new capabilities in computing, communication, sensing, and materials, quantum technologies have the potential to transform various aspects of human life and reshape the global economy making it a technological revolution \cite{sch}. To fully comprehend the potential effect of these technologies and how to prepare for the future, governments and industry leaders must categorise new innovations in the framework of quantum technologies. Policymakers and business leaders can hasten the development of quantum technologies and make sure they are well-positioned to benefit from this revolutionary technology by anticipating potential disruptions, identifying areas for research and development, identifying potential applications, and informing the government on policy decisions.

\section{Addressing quantum readiness}
Although quantum technology is still in its infancy or emerging stages, it is developing quickly. Rapid advancements in quantum technology can expand the global economy by billions of dollars \cite{shelli}. Hence, it is vital to be prepared and adapt to it. In addition to its fast-paced growth, the technology is inherently different from current technologies and will be massively disruptive for most sectors. This technology will likely have a disruptive innovation effect on operations, services, and products, giving companies that exploit it early a significant competitive advantage. In areas where quantum technologies are anticipated to have a big influence \cite{dow}, nations that invest in and embrace quantum technology first will have a competitive edge. Governments that are not quantum-ready run the danger of falling behind in these sectors and may find it difficult to compete with those who have made quantum technology investments  \cite{perrier}. Quantum technologies have the potential to make sophisticated computing, secure communication, and encryption possible \cite{sidhu}. Non quantum-ready entities and organisations may be more susceptible to cyberattacks and other security threats \cite{kilber, qrt}. Quantum computers can defeat current encryption schemes, leaving sensitive data open to theft or abuse. The protection of national interests and citizens will be improved in nations that are quantum-ready. Research advancements in areas such as materials science, drug development \cite{zinner}, and artificial intelligence \cite{krenn} will be made possible by quantum technologies. Researchers must hasten the progress of science and produce important new discoveries by focusing on quantum readiness. Quantum computing can be used to model intricate chemical processes, speeding up the development of novel medicines and materials \cite{bauer}, and has the potential to transform banking by enabling financial institutions to carry out hitherto impractical sophisticated computations and risk analysis \cite{egger}. Global problems including climate change, energy efficiency, and sustainable agriculture might be solved by quantum technologies \cite{berger}. Countries can support sustainable development and the welfare of their citizens by adopting quantum technologies that address these issues \cite{diet}. Quantum computers can potentially be utilised to optimise energy distribution and cut down on waste, while quantum sensors could be used to monitor environmental conditions and increase crop yields \cite{parrish}.

In the near future, quantum readiness will play a key role in deciding if a business/organisation thrives or recedes. For example, \emph{Quantum Key Distribution} (QKD), has the potential to revolutionise the field of cryptography. QKD allows for the secure exchange of encryption keys by exploiting the principles of quantum mechanics. Quantum computers may make current encryption techniques like \emph{RSA} obsolete since they may be able to decrypt data significantly faster than traditional computers. The cybersecurity landscape will drastically change as a result of this disruption \cite{cid}, necessitating the creation and deployment of new cryptographic algorithms that are resistant to quantum computing. In comparison to classical computing, quantum computing can more effectively optimise complex financial models, portfolio management, and risk assessment \cite{deo}. As a result, financial institutions may optimise investment strategies, more precisely assess risks, and more successfully identify fraud, which could potentially have a significant impact on the financial sector.

Since quantum technology can render previous technical knowledge outdated and maintain technological, industrial, economic, and social transformation, it shares many traits with \emph{General-Purpose Technologies} (GPTs). Changes in techno-economic paradigms brought about by GPTs impact all sectors of the economy and continue the long-term process of economic progress in human civilisation. Furthermore, to support an entire and a functional quantum ecosystem built on solid physical infrastructures, highly skilled human resources, and suitable technological systems, the evolution of this technology requires time, research and development investments, carefully chosen research policies, and key strategic decisions of governments and industry. Therefore, quantum readiness comes in with a first-mover advantage and the power to shape the direction of the technology. For instance, organisations in the pharmaceutical and materials sectors that use quantum computing to identify new drugs or design new materials might accelerate research and development efforts. These businesses can obtain a first-mover advantage by discovering potential drug candidates or novel materials more rapidly and accurately than rivals, that use classical computing approaches, enabling them to sell new products quickly and possibly secure patents earlier allowing them to thrive \cite{cao}.

For organisations to be competitive, safe, and economically prosperous in the future, addressing quantum readiness is essential. Countries can capitalise on this game-changing technology and promote science and sustainable development by investing in quantum technologies and building a strong quantum ecosystem. Although there are many obstacles to creating and using quantum technology, the advantages are too enormous to ignore. Organisations can make sure they are ready to profit from this innovative new technology by collaborating to address the issue of quantum readiness.

\section{Diffusion of Emerging Technology}
Building on our understanding of quantum technologies' potential impact, it becomes imperative to delve into the concept of quantum readiness as discussed in the previous sections. But as we navigate this pioneering landscape, the importance of having tools to gauge the progress of quantum technologies cannot be overstated. Such instruments would allow us to systematically assess the evolution of these technologies, drawing from various conceptual models and metrics. They can serve as compasses for interested entities, whether they are research organisations, industry stakeholders, or policymakers. With a tangible sense of where we stand on the quantum technology trajectory, we can better identify the risks involved, thus enabling more strategic planning. The subsequent examination of quantum readiness is therefore not only a scholarly exercise but a crucial step in mobilising the global quantum ecosystem.

Radical innovation models refer to frameworks used to explain how new and emerging, transformative technologies or business models emerge and disrupt existing industries or markets. It is still early to see the full impact of quantum technologies, but many (including the authors) expect them to be a highly transformative and disruptive set of technologies. Therefore, we argue that investigating and thinking about quantum technologies via radical innovation models is a suitable theoretical framework. One such model is the \emph{Disruptive Innovation Model} developed by Clayton Christensen, which explains how new technologies or business models can initially serve niche markets before eventually disrupting and overtaking established ones \cite{Christensen1997}.

Another potentially relevant concept is pervasive innovation, which refers to the idea that innovation can occur across all aspects of society and in all sectors of the economy, leading to profound changes in how people live, work, and interact with one another \cite{markides2013}. This can be thought of in relation to the idea of the techno-economic revolution, a concept which refers to the profound economic and societal changes that can result from the emergence of new technologies or industries \cite{perez2010}. These revolutions often involve the displacement of existing industries, the creation of new ones, and shifts in the distribution of wealth and power. Examples of techno-economic revolutions include the industrial revolution, the information revolution, and the ongoing transition to a digital economy.

Techno-economic paradigms should also be mentioned here in its relation to techno-economic revolutions. These are overarching frameworks that guide technological innovation and economic growth in a particular period. They represent the dominant technological and economic structures, processes, and institutions that shape innovation and growth in a given era \cite{dosi1998}. Techno-economic paradigms are characterised by a set of shared beliefs, values, and practices that define the boundaries of what is considered technically feasible and economically viable. It is debatable whether quantum technologies should be viewed as a potential new techno-economic paradigm or as a subset of digital transformation. However, the analysis and recommendations developed in this paper are irrespective of either case. Quantum technologies, either by themselves or as a part of the constellation of the technologies enabling the transition to the next techno-economic paradigm, will play an important role in the following decade.

Both radical or pervasive innovation, techno-economic revolutions and paradigms are important concepts to consider in the context of the period we are experiencing in relation to the second quantum revolution and associated quantum technologies. By acknowledging the potential for innovation to arise from unexpected sources and recognising the profound economic and societal changes that can result from new technologies, organisations can better prepare for and navigate the challenges and opportunities of the innovation process. In this regard, one can turn to transitional studies, which focus on how organisations or industries can manage and navigate transitions to new technologies or business models. These studies examine the challenges, opportunities, and strategies involved in moving from current practices to new ones \cite{geels2017}.

Quantum readiness, in this context, refers to the ability of organisations to adopt and leverage quantum technologies for competitive advantage, especially during the transitional period where profound economic and societal changes may occur due to the radical and pervasive nature of quantum technologies. This set of technologies, including quantum computing and quantum cryptography, has the potential to revolutionise many vertical industries such as finance, healthcare, and logistics, but it also requires significant investments in infrastructure, expertise, and cultural change. Therefore, this process of getting ready for and enabling the second quantum revolution can be thought of as a highly entangled process.

The application of radical innovation models and transitional studies can help organisations prepare for and navigate the transition to quantum readiness. For example, the Disruptive Innovation Model can provide insights into potential market disruptions and the emergence of new business models in the quantum space, while transitional studies can help organisations identify and manage the barriers and the enablers of quantum adoption, such as funding, talent, and regulatory frameworks.

In summary, radical innovation models and transitional studies are valuable tools for organisations seeking to become quantum-ready. By understanding the dynamics of radical innovation and managing the transition to new technologies, organisations can position themselves for success in the emerging quantum landscape. Three common tools that are widely used in this literature are the well-known \emph{S-curve} \cite{murat}, the \emph{Technology Readiness Levels} (TRLs) framework \cite{mankins, conrow} and \emph{Technology Commercial Readiness Levels} (TCRLs) framework \cite{ani}.

\subsection{S-curve analysis}
The S-curve \cite{sawag} diffusion model is a useful tool for predicting and understanding the adoption and diffusion of new technologies, including quantum technology. It can be used to predict the rate of adoption of quantum technology and identify potential opportunities and challenges. For example, companies can use the S-curve model to identify when quantum technology is likely to reach the growth phase and plan accordingly \cite{maria,li}. This can include investing in research and development, building partnerships, and developing new products and services.

\begin{figure}[h]
    \centering
    \includegraphics[scale=1]{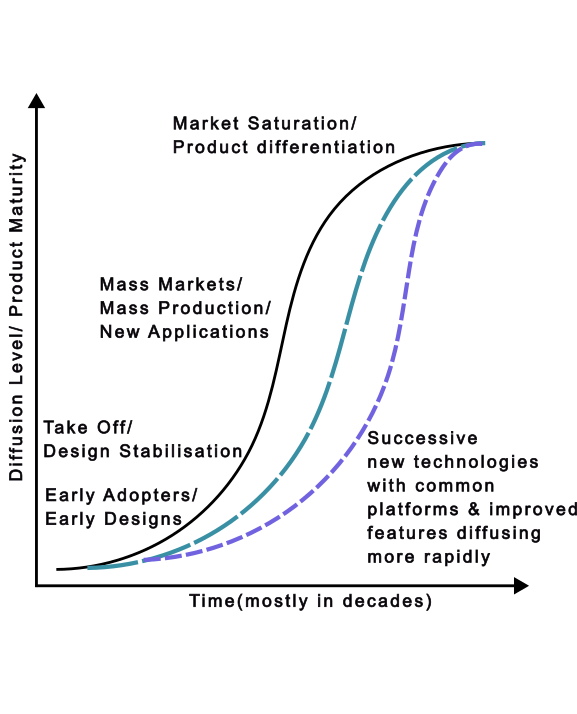}
    \vspace{-1.5cm}
    \caption{The S-curve analysis of emerging technologies}
     
    \label{Fig.2}
\end{figure}

As the growth of many innovations follows the fundamental pattern, the uptake of innovation is frequently represented in these terms in diffusion analysis \cite{vernon}. Slow adoption is followed by a period of high growth, which is then followed by slower growth as the majority of potential adopters have already adopted the new product (Figure ~\ref{Fig.2}) \cite{cawson}.
S-curve diffusion assessments frequently highlight the differences between early and late users. Early adopters tend to be wealthier, more educated, connected to mass media, etc. on an individual level. We may also assume that there is a pattern of diffusion across business sectors, with some being early adopters and some being followers. This is true if we look at the dissemination of significant new technologies that can be used for a variety of purposes. Early adopters, for instance, frequently work in high-tech industries and are presumably more closely connected to the fields where the technology was initially produced and/or used.
The analysis can be approached through the lens of the industry life cycle and product life cycle \cite{utb}. The basic idea behind the industry life cycle approach is that sectors are likely to \emph{mature} over time. Their operations become conventional, their goods become more standardised, and the production tools they need to use become more affordable or easier to use. They rely less on highly skilled labour.

The product life cycle approach makes additional comments that are especially relevant when taking the dynamics of new technologies into account. This strategy incorporates the findings of several innovation studies as well as the concepts of diffusion and industry cycles. Here, the essence of the technology itself is more important than how the market or industry develops. According to the product life cycle concept, early versions of inventions are frequent, even when technologically advanced, relatively simple and primitive in comparison to their later versions. Not merely because the marketplaces are still in their infancy, but because the early generations of innovations appeal to very few customers. There is a lack of understanding, on the user side of their existing and future capabilities, and on the supplier side, of precisely what skills will be valued and how they may be employed. It also prompts the idea that there could be weak connections between inventors and numerous prospective consumers and use cases of the technology in the early stages of product creation and distribution. Complementary goods and services are either unavailable or not generally accessible in the early stages of development. Comparing the essential components to subsequent versions of comparable products, the key products are likely to be more expensive and less reliable. Early versions of goods are likely to need significant technical expertise for their manufacturing and usage, but later versions may make use of these abilities as a common practice. If the products succeed, people will become more aware of them, invest more in them, and acquire more skills using them and their markets will grow. The technology itself will evolve, which is essential. The product has undergone redesigning to make it more durable, user-friendly, and capable of more effective manufacturing. New entrants, who could bring fresh concepts for innovation, join the early suppliers, who have shown the potential for a sizable new market. Several innovation approaches frequently compete with one another, with the winner establishing the design paradigm to which all others should adhere. If the first providers are small firms, big businesses are liable to substantially change the nature of the competition in the market possibly by acquiring the smaller firms. Large enterprises with greater marketing capabilities have the capability to boost the diffusion process and the creation of a dominant design. Later, as the market expands, the emphasis on innovation often shifts away from fundamental product/technology innovation and towards providing enhanced quality before shifting to process innovation that is scalable and economical. With supplementary goods and services, enabling technologies, increased functionality, and greater adaptability, the product becomes more user-friendly and requires fewer high-level skills to utilise. There can be a shift from technology-push to demand-pull. Successful inventions have an extended development phase once they are introduced to the market, not just while they are pre-commercial prototypes in research and development institutions, but as providers discover what consumers want. Users, on the other hand, acquire knowledge about the product and successful usage techniques.

When it comes to quantum technology it gets tricky, as it itself branches into several technologies, including quantum computation, sensing, communication, cryptography, and more. Each of these individual technologies stands at different stages. Furthermore, some of the technologies involve similar architectures and platforms, although currently there is no clear winner for which platform will be adopted in general or within its branches.
Therefore, it is worthwhile to look at quantum technology in general to assess the current scenario. The adoption rate of quantum technology is examined over time, and it is compared to the adoption rates of other technologies, in the S-curve analysis. Quantum technology adoption also entails evaluating the elements influencing or impeding its uptake as well as projecting its possible future uptake which we call indicators. By considering indicators such as research publications and patents, investment, commercial products and services, talent pool, technical advancements, industry partnerships, and government policies, we can evaluate the current status and potential future of quantum technology \cite{rueda}. This understanding is essential for determining the quantum readiness level, which is critical for realising the full potential of quantum technology.

\begin{figure}[h]
    \centering
    \includegraphics[scale=0.7]{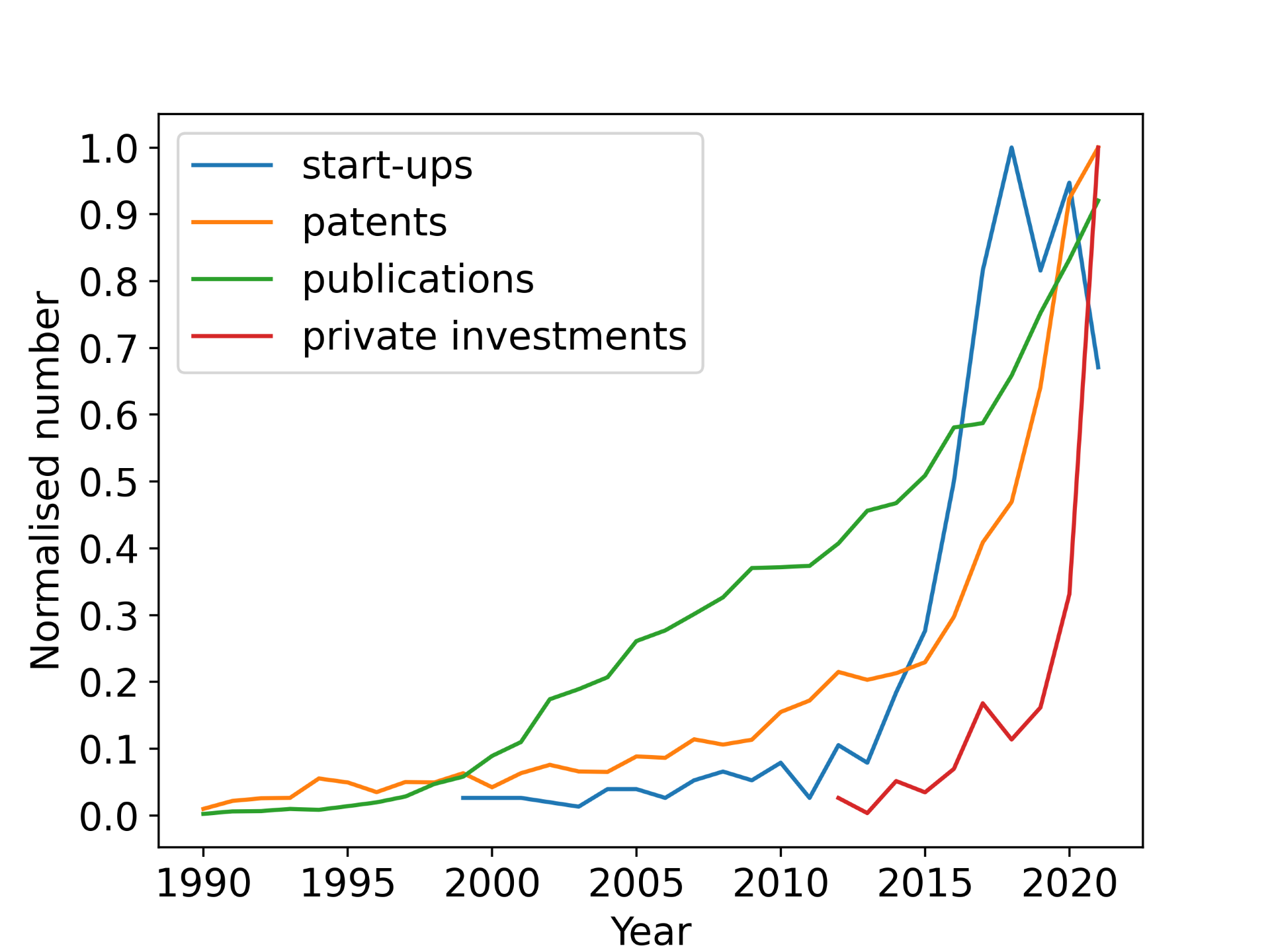}
    
    \vspace{0.15cm}
    \caption{Various quantum readiness indicators over time for quantum technologies}
     
    \label{Fig.3}
\end{figure}

When we say adoption of a technology, we refer to the extent to which the technology is being used or integrated within a sector or an ecosystem reflecting acceptance of the technology by consumers, industries, or society as a whole. Determining and choosing the Key Performance Indicators (KPIs) can help us infer the level of adoption by showing where the technology is in its life cycle. For example, an increase in research publications over time suggests that the technology is gaining interest in the academic community. An increase in investment in a particular technology indicates growing confidence in its commercial potential. Start-ups represent the presence of commercial products and services which is a clear sign that a technology is being adopted. It is essential to recognise that these indicators do not occur in isolation, but rather interact and influence each other. For example, an increase in research publications can lead to more patents, which can then attract more investment. This cascading effect can accelerate the technology's movement along the S-curve, hastening its adoption. Plotting these indicators over time can give us a visual representation of the technology's progression along the S-curve of adoption. It provides a multi-dimensional perspective of the technology's adoption status, with each indicator representing a different aspect of the technology's life cycle.

To see a general trend of the growth and diffusion of quantum technologies in general, we collected data for various indicators over time and plotted it on the same time scale with the total number of each indicator normalised to one (our code reads in data from four CSV files containing the number of startups, patents, publications, and private investments for each year. It then normalises the data so that the maximum value for each category is equal to 1 for a better visualisation to compare these trends) in Figure ~\ref{Fig.3}. For the collection of data on publications we used bibliometric tools such as the \emph{Web of Science} (WoS) database and search query as done in \cite{Seskir2021}.
For patents, we used the data recently published in the paper \cite{seskir_patent_2023}. For start-ups data, we used the numbers from \cite{Seskir2022} and the database maintained and updated by QURECA.
For private investments, we used the data from the Quantum Technology Investment Update 2022 report published by The Quantum Insider \cite{matt}. When we plotted all these data against years, we see a general trend of take-off of quantum technologies, indicating that now is the perfect time to become early adopters of quantum technology. A conjoint peak in start-ups, patents, publications, and private investments in a particular technology can be seen as a strong indicator that the technology is ripe for early adoption. When these indicators peak simultaneously, it signifies a strong momentum behind the technology. This synchronisation is often seen when the technology is transitioning from the innovation stage to the early adoption stage in the technology adoption lifecycle. These factors suggest a good time for early adoption, they do not eliminate the inherent risks associated with emerging technologies. However, emerging technologies are inherently risky and can fail for a variety of reasons, including technical challenges, market acceptance issues, regulatory problems, and many others. An alignment of these indicators does suggest that the technology is maturing and moving towards broader market adoption. This can be an ideal time for organisations that are comfortable with a certain level of risk to become early adopters. By doing so, organisations may benefit from leveraging the technology before it becomes widespread, while also having a chance to shape the technology's development and application in their industry. Being an early adopter of quantum technologies can provide organisations with a range of benefits. It can help them to stay ahead of the curve giving them a competitive advantage, improving productivity, innovating new products and services, and shaping the direction of the technology. Although we could see some dips in 2023, it could be an implication of various factors away from normal conditions such as the Covid-19 pandemic and more.

Forecasters must be aware that there are sometimes variations in the norm. Even though S-curves typically offer quite good fits to empirical data, in reality, they are frequently disrupted by other factors such as wars and economic downturns. Innovations can also replace one another before a certain innovation is adopted by all its prospective users, and a rival technology may supplant it.

\subsection{Technology readiness levels for quantum technologies}
\emph{Technology Readiness Levels} (TRLs) are a method of measuring the maturity of a technology by determining its level of development, testing and integration \cite{ole}. The TRLs provide a common framework for evaluating and communicating the maturity of a technology and can be used to identify areas where further research and development are needed. For quantum technologies, it makes more sense to use TRLs to assess the readiness of individual branches or at an individual product level. It is crucial to identify the quantum technology to be assessed and its intended application(s). This will help in identifying the specific requirements for the technology and the performance metrics needed to assess its TRL. Then one must determine the TRL criteria for the quantum technology under assessment. This can include factors such as the level of technical feasibility, the maturity of the design, the level of testing and validation, and the readiness for deployment. Evaluating the technology against the TRL criteria by analysing the performance data from experimental studies, simulations, and tests, and comparing the results helps assign a TRL value. This should reflect the technology's current stage of development and the level of maturity it has achieved. It also reflects the level of confidence with which the technology is ready for deployment in the market. Based on this TRL, one can then develop a roadmap for the technology. This involves identifying the key development milestones needed to reach the next TRL and the associated resources and investments required.

TRLs are typically defined on a scale of 1 to 9 \cite{h2020,trl0}. The same can be adopted for quantum technologies with the addition of ethics protocols for the branch of quantum technology in consideration. The first four levels usually address the most fundamental technical research involving, mostly, laboratory results given the sort of research, technological advancement, and innovation being addressed. From levels TRL 5 through TRL 6, technological development would then proceed until the first prototype or demonstrator is obtained. Projects involving technological innovation would fall between TRL 7 and TRL 9, as this type of innovation necessitates the launch of a new product or service onto the market, which involves passing the necessary tests, certifications, ethics, and approvals. These stages entail deployment or extensive implementation. Assessing the readiness of new technology products is important for understanding their maturity and potential for commercialisation. 

\begin{figure}[h]
    \centering
    \includegraphics[scale=0.52]{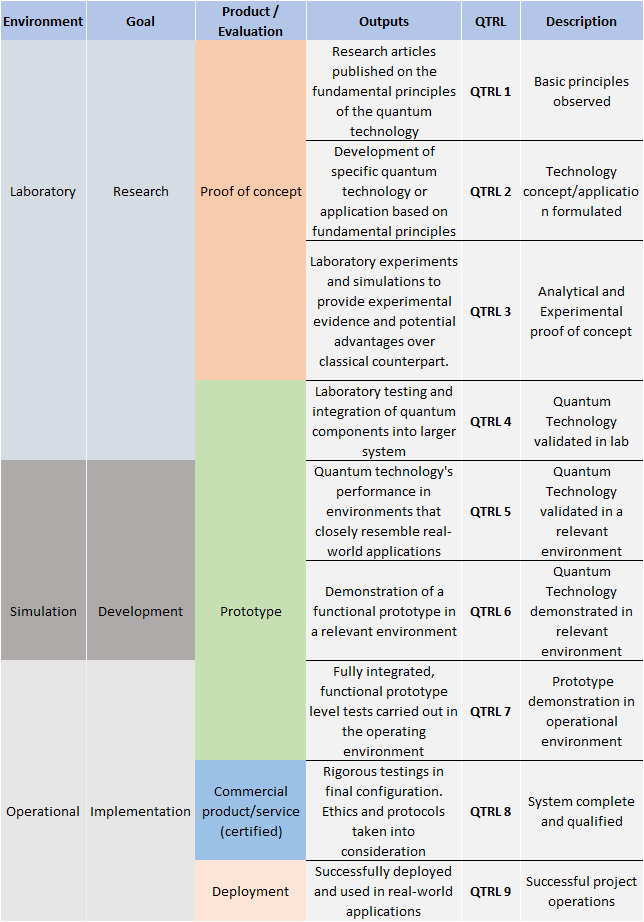}
    \vspace{0.15cm}
    \caption{\emph{Quantum Technology Readiness Levels} (QTRLs) description, representation and expected outcomes}
     
    \label{Fig.4}
\end{figure}

We have formulated standard Quantum Technology Readiness Levels(QTRLs) taking inspiration from the ones set for general TRLs and TRLs for artificial intelligence technology to help assess a quantum technology in more accurately as shown in Figure ~\ref{Fig.4}. The description and expected outcomes from each of these levels are described below:
\begin{enumerate}
  \item QTRL 1 (Basic principles observed): Research into the basic principles of quantum phenomena is at present being conducted. This might involve researching various phenomena such as quantum superposition, entanglement, and coherence as well as creating theoretical equations and models to explain these concepts.
  \item QTRL 2 (Technology concept/application formulated): Based on the concepts observed in QTRL 1, researchers develop a particular quantum technology concept or application at this phase. This entails the formation of a distinct technical vision and goals and may include concepts for quantum computing, communication, or sensing applications.
  \item QTRL 3 (Analytical and experimental proof of concept): Researchers working at this level offer analytical and experimental evidence that the technological concept or application is feasible. This typically involves laboratory experiments and simulations to validate the technology's functionality, performance, and potential advantages over classical technologies.
  \item QTRL 4 (Quantum technology validated in lab): Quantum technology systems or components are developed and assessed at QTRL 4 in a controlled lab setting. This stage focuses on the integration of individual quantum components into a larger system and the testing of their performance, reliability, and scalability.
  \item QTRL 5 (Quantum technology validated in relevant environment): In this phase, researchers validate the system's performance in environments that closely resemble real-world applications. This might involve testing quantum communication systems over long distances, assessing the performance of quantum sensors in realistic settings, or benchmarking quantum algorithms on prototype quantum processors.
  \item QTRL 6 (Quantum technology demonstrated in relevant environment): QTRL 6 requires a demonstration of a fully functional system or subsystem within a relevant environment. This could include a quantum communication network connecting multiple users, a quantum sensor deployed in the field, or a quantum processor executing a specific algorithm to solve a real-world problem.
  \item QTRL 7 (Prototype demonstration in operational environment): At this level, a fully integrated system prototype is demonstrated in an operational real-world environment. This may involve field testing of a quantum communication system for secure data transmission, deploying a quantum sensor for environmental monitoring, or running a quantum computer to solve complex optimisation problems.
  \item QTRL 8 (System complete and qualified): QTRL 8 is achieved when the system has been completed and qualified through rigorous testing and demonstration. This includes achieving all performance requirements, addressing any identified issues or limitations, and demonstrating a high level of reliability and robustness. Ethics and other protocols are also taken into consideration and checked.
  \item QTRL 9 (Successful project operations): When the quantum technology has been successfully deployed and used in practical real-world applications or missions, the final QTRL has been reached. This may entail the widespread use of quantum sensors in a variety of businesses, the commercial availability of quantum computing services, or the use of quantum communication technologies for secure data transmission.
 
 \end{enumerate}

\begin{figure}[h]
    \centering
    \includegraphics[scale=0.46]{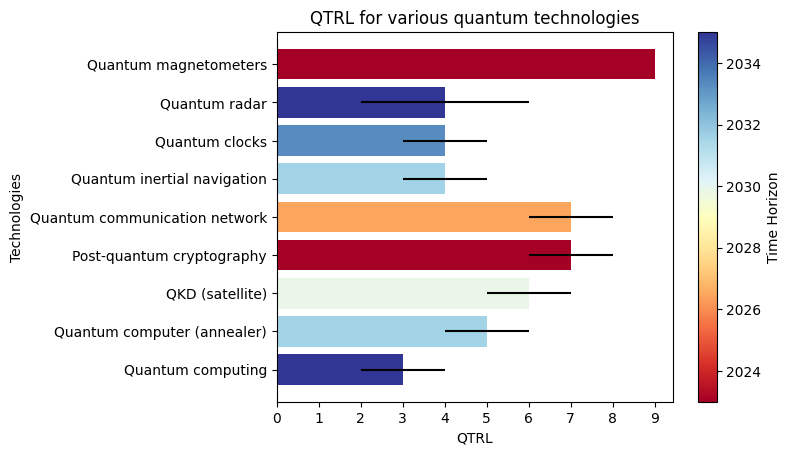}
    \vspace{-0.5cm}
    \caption{QTRLs and time horizon expectations (with error bars) for several quantum technologies}
     
    \label{Fig.5}
\end{figure}

One way to assess the readiness of a new technology product is to use a readiness-vs-generality chart \cite{jose}. The same can be used if one wants to evaluate the readiness of a specific quantum technology-based product and not a branch in general. One must identify the different layers of the capability of the product in relevant environments and then compare it against the TRL of each layer of capability. 
 For instance, Forschungszentum Jülich defined a set of TRLs for quantum computing \cite{fzj}. The theoretical foundation for quantum computing (annealing) is developed when quantum computing technology is at QTRL 1. Once the fundamental device principles have been investigated and applications or technologically pertinent algorithms have been developed, the technology reaches QTRL 2. The fundamental components of quantum computing systems, physical qubits that have been fabricated imperfectly, are at QTRL 3. Laboratory tests are then designed to verify the theoretical predictions of qubit characteristics and then proceed to the fabrication of multiple qubit systems with the classical control unit in QTRL 4 stage. Technology for QTRL 5 quantum computing consists of parts that are incorporated into a tiny quantum processor without error
correction. At QTRL 6, there are components assembled into a miniature error-correcting quantum processor. They are rigorously tested with various quantum algorithms and benchmarked. In the QTRL 7 stage, the prototype is tested for solving small, but relevant use-case problems. Once the prototype is made scalable and qualifies for all necessary tests, it advances to QTRL 8 stage. Finally, when the prototype quantum computer exceeds the power of the classical computers in specific problems, it is labelled as QTRL 9. This acts as a critical tool for them to assess their current quantum readiness level and helps in strategising a roadmap for their prototype and planning future steps of action and set goals. Hence, it is of utmost importance that QTRL assessment is conducted by qualified experts with a deep understanding of the technology and its development. It is also important to set standardised QTRL levels in each branch of quantum technology so as to make meaningful comparisons with other competing organisations/platforms.
One of the major highlights of Quantum.Tech, a conference held in 2022 in Boston, Massachusets, USA was assessing the TRLs for various branches of quantum technology. In comparison to quantum sensing and quantum computing, quantum communication, such as QKD, is viewed as the more developed domain (QTRL 7) for many future implementations while quantum computing was seen to be at QTRL 3 \cite{andi}. Another report in 2021 to assess quantum readiness for military applications provided a detailed report on various TRLs for different quantum technologies \cite{krelina}. Taking data from all these reports and roadmaps, we provide an assessment of current TRLs for various quantum technologies and a time horizon indicating the expected time it would take to achieve a QTRL of 9 for these quantum technologies (Figure ~\ref{Fig.5}).  

We can clearly see that quantum technologies are at different QTRLs. By considering diverse applications and deployment platforms, the QTRL variance and time horizon assumptions become much more complicated. Building on these reports provides a snapshot of trends and should reduce sampling bias. In conclusion, the QTRLs of different branches of quantum technologies vary widely, with quantum computing being in the earliest stages of development and quantum communication and cryptography being more mature. Significant progress has been made in recent years in all branches of quantum technologies, but challenges remain in scaling the technology to meet the requirements of practical applications. It is crucial to keep a track of the QTRL levels of these branches in order to stay ahead of the curve and be quantum-ready.

\begin{figure}[h]
    \centering
    \includegraphics[scale=0.25]{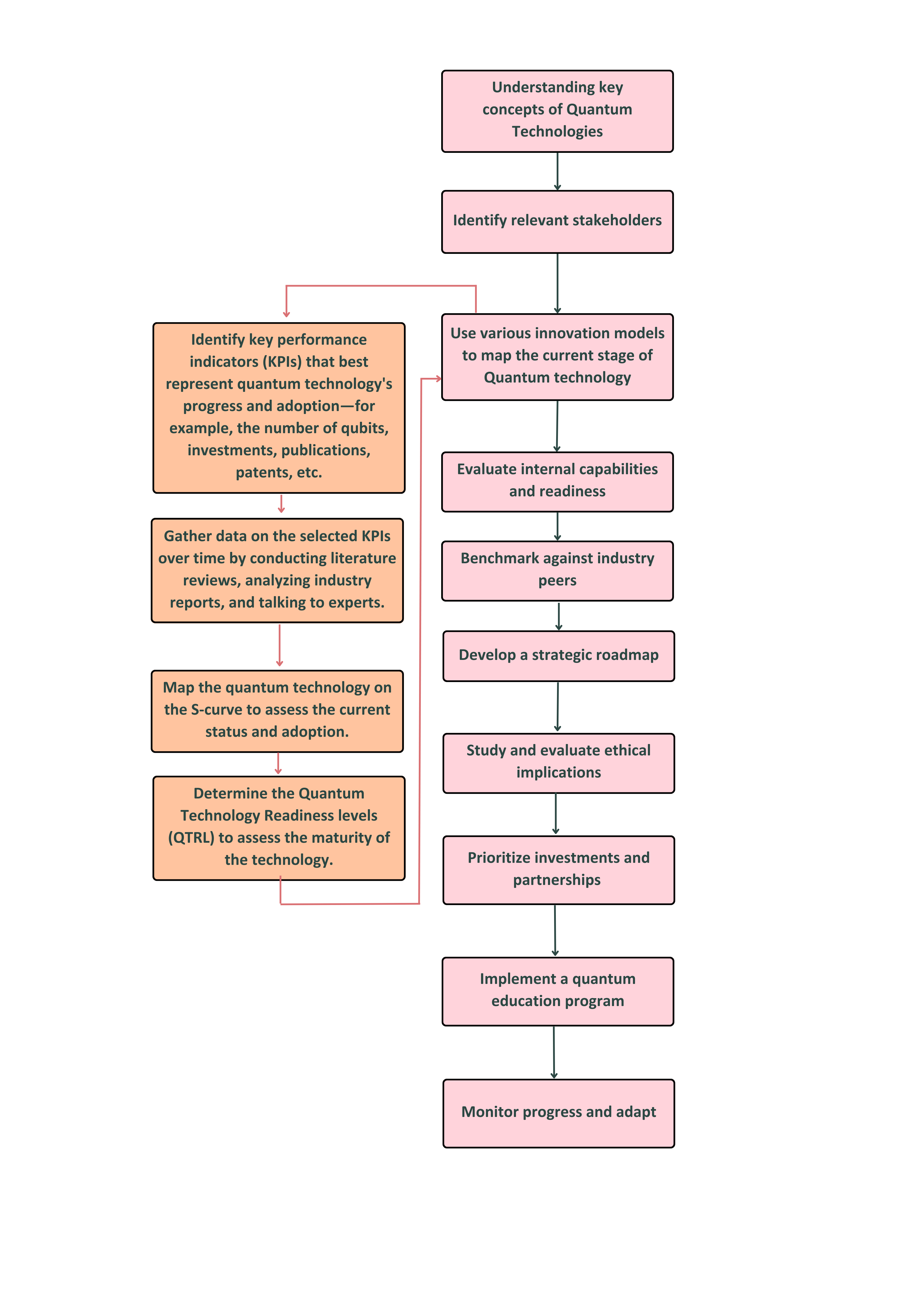}
    \vspace{-1.5 cm}
    \caption{Flow chart describing the steps to follow to be quantum-ready}
     
    \label{Fig.6}
\end{figure}

With all these tools and information, we have devised a roadmap or a flowchart with detailed steps to assess quantum readiness (Figure ~\ref{Fig.6}). It will help businesses and organisations to assess their quantum readiness or to develop a strategy to be quantum-ready. It can be summarised as follows:
\begin{enumerate}
    \item Understanding key concepts of Quantum Technologies: Gain a basic understanding of quantum physics principles, such as superposition, entanglement, qubits, and sensors.
    \item Identify relevant stakeholders: Assemble a team of stakeholders from different departments such as research and development and executive leadership. This team will be responsible for driving the assessment process and implementing the findings.
    \item Use various innovation models to map the current stage of quantum technology: Identify and gather data over time for various KPIs (Key Performance Indicators)and map them on the S-curve, evaluate TRL to assess maturity of quantum technology.
    \item Evaluate internal capabilities and readiness: Evaluate your organisation's current capabilities and readiness to adopt quantum technology. Identify any gaps in knowledge, expertise, or resources that need to be addressed. 
    \item Benchmark against industry peers: Analyse your competitors and industry leaders to understand their quantum readiness and strategies. Identify best practices. 
    \item Develop a strategic roadmap: Create a strategic roadmap outlining your organisation's plan to adopt and implement quantum technology. This should include short-term and long-term objectives, as well as a timeline for implementation.
    \item Study and evaluate ethical implications: Explore, assess, and adapt ethical and societal responsible use of quantum technology. 
    \item Prioritise investments and partnerships: Determine the resources required to achieve your quantum technology readiness goals, such as investments in hardware, software, and talent. Prioritise investments based on potential impact and return on investment. Identify potential partners, such as universities, research institutions, and technology providers, to collaborate on quantum technology initiatives.
    \item Implement a quantum education program: Develop and implement a comprehensive quantum education program for your organisation. This may include training sessions, workshops, and seminars to ensure all relevant stakeholders are informed about the latest developments in quantum technology.
    \item Monitor progress and adapt: Regularly review and update your quantum technology readiness assessment and strategy based on new developments in the field. Continuously evaluate your organisation's progress and adjust as necessary to stay on track with your quantum technology goals. Publish these assessment reports. 
    
    \end{enumerate}
\subsection{Commercial Readiness of quantum technologies}
While Technology Readiness Levels (TRLs) provide an assessment of a quantum technology's technical maturity, they do not account for its market readiness or commercial viability. Quantum Commercial Readiness Levels (QCRLs) offer a strategic framework for determining the market readiness of quantum technologies. They examine the commercial aspects of these technologies, focusing on their potential to generate revenue, their fit within the market, and their capacity to navigate the regulatory environment \cite{greig, ani}. We provide a framework to assess it as follows:
\begin{enumerate}
 \item QCRL 1 (Problem-Solution Fit): At QCRL 1, the primary goal is to demonstrate that a quantum technology can offer a viable solution to an identified problem. It is an ideation phase, where researchers and developers are confirming that their quantum-based solution aligns with a particular need or problem in the market involving rigorous problem definition and in-depth market research crucial to that time. The identified problem may be broad, encompassing challenges in various sectors like cybersecurity, material sciences, logistics, or drug discovery, among others. It is essential to establish a preliminary hypothesis for how the technology can address these issues and identify potential customer segments who might benefit from the solution.
 \item QCRL 2 (Market-Solution Fit): The focus now is on validating the proposed solution within the potential market. This involves in-depth market research to understand the needs and preferences of potential customers, evaluating the competitive landscape, and performing a preliminary assessment of the regulatory landscape. The aim is to fine-tune the solution to fit the market needs better and establish initial relationships with potential customers or partners. It is at this stage that a technology might pivot or refine its focus, based on market feedback and competition.
 \item QCRL 3 (Minimal Viable Product (MVP) Development): Upon reaching QCRL 3, the technology has progressed to a point where a minimal viable product (MVP) can be developed. This MVP encapsulates the essential functionality that addresses the identified problem, serving as a tangible representation of the technology’s potential in the real world. It is used to solicit feedback from early users, allowing for further refinement of the product based on actual user experience. This stage involves significant interaction with potential customers, taking in their feedback, and adjusting the product's design and function accordingly.
 \item QCRL 4 (Product-Market Fit): A full-fledged, commercially viable product has been developed now based on feedback from the MVP stage. This stage validates the product’s market fit, demonstrating that there is a market demand for the product, and that the product satisfies that demand adequately by rigorous benchmarking. Business models are tested and revenue generation strategies are defined. Early sales to innovators or early adopters have been achieved, providing evidence of the product’s market fit.
 \item QCRL 5 (Scaling and Growth): In the final stage of commercial readiness, where the technology is ready for full commercialisation and scaling. The focus is on growing sales, expanding into new markets, and optimising the product based on customer feedback and market dynamics. The production and distribution processes have been solidified, the supply chain has been identified, there is a clear marketing strategy, and any pertinent regulatory hurdles have been navigated by this time.
\end{enumerate}
Quantum technologies often require significant investment and time for development. By assessing the QCRLs, investors and developers can better understand the commercial risks associated with a particular technology and make informed decisions about resource allocation. It helps in identifying potential market barriers early, which can be addressed proactively, reducing the likelihood of costly failures down the line. Furthermore, the QCRLs facilitate regulatory planning, helping companies anticipate potential regulatory hurdles and strategise accordingly. Finally, by showcasing a technology's commercial readiness, QCRLs can instill confidence in potential customers, accelerating the adoption and acceptance of quantum technologies in the market. In essence, QCRLs provide a pragmatic framework that is essential for strategic decision-making and for driving the successful diffusion of quantum technologies across various sectors.
\\
Mapping QCRLs to QTRLs is an essential step in the journey of quantum technology development and commercialisation. This process brings a necessary cohesion between the technical progression and market readiness of a quantum technology, offering a comprehensive view of its trajectory from laboratory to market. The initial QTRLs (1-3), where technologies are in the research and development phase, correspond to QCRL 1, where the problem-solution fit is being explored. As the technology progresses to the prototype phase (QTRLs 4-5), it aligns with QCRLs 2 and 3, where market validation and MVP development are occurring.
At the demonstration stage of the QTRLs (6-7), technologies typically reach QCRL 4, where the product-market fit is established, and early sales are achieved. At the higher QTRLs (8-9), where the technology is ready for deployment or already in use, they align with QCRL 5, indicating that the technology is ready for full commercialisation and scaling.
However, it is essential to understand that once a technology reaches QTRL 9, there can still be
significant barriers to widespread adoption, including cost, regulatory hurdles, workforce
training, and more.
A common issue in technology development is the "valley of death," a phase where a technology, despite being technically sound, fails to reach the market due to commercial hurdles \cite{nemet}. By mapping QCRLs to QTRLs, stakeholders can better navigate this valley, as they gain insights not only about the technology's technical maturity but also its commercial viability at each stage of development. Such mapping supports informed decision-making and risk management. By understanding where technology sits in both the technical and commercial spectra, developers, investors, and policymakers can align their strategies and resources more effectively. They can anticipate potential technical and market barriers, prepare for regulatory requirements, validate their value proposition and business model, and determine optimal market entry strategies.
\section{Quantum readiness for different stakeholders}

As quantum technology moves from the realm of theoretical research into the field of practical applications, it is essential to understand and foster a robust and effective quantum ecosystem. This ecosystem refers to the interconnected network of stakeholders, including researchers, industry practitioners, policymakers, educators, and investors, all contributing towards the development, adoption, and regulation of quantum technologies. A robust and effective ecosystem is the one that ensures a fluid exchange of knowledge and resources among its participants, facilitating continuous innovation, standardisation, and commercialisation of quantum technologies. It thrives on collaboration and mutual support, fostering perennial growth amidst market fluctuations and technological uncertainties. Participants in a robust quantum ecosystem are expected to assume distinct yet interconnected roles and responsibilities. Researchers, for instance, are tasked with advancing the scientific frontier of quantum technologies, while educators are responsible for cultivating a future-ready workforce equipped with quantum literacy. Industry practitioners transform theoretical research into market-ready solutions, continuously pushing the practical boundaries of quantum applications. Policymakers establish regulatory frameworks, balancing innovation, security, and ethical considerations. Lastly, investors provide the necessary financial resources, driving the ecosystem from early-stage research to commercially viable products. All these roles contribute to the vibrancy and resilience of the quantum ecosystem, each indispensable for fostering quantum readiness across different sectors. Hence, it becomes cardinal to understand the role of these stakeholders in quantum readiness.
In the previous sections, we discussed the interpretation of quantum readiness and various models to assess current quantum readiness by using different models. Next, we describe in detail various perspective scenarios in which the term \emph{quantum readiness} is predominantly used and of relevance for key stakeholders in the quantum ecosystem. We discuss what it means to be quantum-ready in each of these scenarios and its relevant indicators. 

\subsection{Government}

Common themes for a government to be quantum-ready are to have the necessary infrastructure, expertise, and resources to develop, implement, and use quantum technologies. The following section outlines structures found across nations developing their quantum-readiness, namely strategies and roadmaps, ecosystem and supply chain, and ethics and policy. 

National quantum strategies are a common framework for a government to identify areas where quantum technologies could have the most significant impact, such as in finance, healthcare, or national security. By focusing its investments \cite{qureca2023} in these areas, a government can develop the necessary infrastructure and expertise to become a leader in quantum technologies. A roadmap is essential to forecast the implementation of quantum technologies for each of the following points:

\begin{enumerate}

    \item	Research and development: Investment in basic research is essential for advancing the state of quantum technologies. This includes funding for universities, research institutes, and national labs to conduct basic research on quantum materials, devices, and systems.
    \item	Education and training: Investment in education and training programs is critical to develop the workforce with the necessary skills and knowledge to design, develop, and implement quantum technologies.
    \item Infrastructure: Investment in premises is essential to support the infrastructure and deployment of quantum technologies in specific industries or applications, including the development of testing and measurement facilities, as well as the establishment of quantum computing centres.

\end{enumerate}
    
The development of coordinated local ecosystems and the understanding of supply chains are essential for quantum readiness. A secretariat sitting within a branch of the government is helpful to achieve several actions across the ecosystem, namely collaboration with stakeholders, coordination of resources, monitoring of implementation, engagement with international partners, public outreach, as well as regular updates and reviews. Such a collaborative approach is best practice to share expertise, resources, and to accelerate the development and adoption of quantum technologies. Such partnerships between the government, companies, research centres, and supporting organisations can ensure that the supply chain for quantum technologies is secure and resilient. Strategic development of domestic capabilities to produce quantum components, as well as their implementation in global supply chains, can accelerate commercialisation and improve supply chain resiliency.
An outcome of a strategy is to help a government address the ethical and policy issues that can arise with the development and use of quantum technologies. This could include issues related to data privacy, cybersecurity, and the impact of quantum technologies on society. The implementation of standards and certifications for quantum components can ensure that they meet certain security and performance requirements, thus further enhancing the security and reliability of the supply chain. Standards and certifications, in particular, can help build public confidence by providing a benchmark against which the performance and security of quantum technologies can be measured. This can help alleviate concerns about the reliability and safety of quantum technologies, especially for applications such as healthcare, finance, and national security.

\subsection{Industry}
Industries, in the conversation about quantum readiness, play a pivotal role in the development and maturation of a quantum ecosystem. They should be key parts of strategies and influence governments to make structural and policy changes to attract other users. For a specific company, the journey to be quantum-ready could differ, mainly depending on the size of the company, or how the company embraces new technologies. Overall, the only way to become quantum-ready is by a step-by-step approach:

\begin{enumerate}
\item	Firstly, understanding quantum is critical. For decision makers, an awareness of the foundations and basic principles of the technology in general terms will be sufficient, whereas more technical knowledge will be useful for technical professionals. Being aware of how the technology develops, as well as starting to get involved in the ecosystem is very important for any organisation. To start, a list of available resources can be found in \cite{Kaur}.
\item	Secondly, the understanding, at a high level, of how quantum technologies can be impacting the business is needed. At this point, most organisations identify an individual or a group of individuals who will take the lead on quantum readiness for the company. Building skills and awareness at this level is crucial.
\item	At a later stage, companies will identify specific use cases and start with a proof of concept or a demonstrator, working with quantum startups. This early experimentation will bring a competitive advantage to the business \cite{sodhi}. 
As quantum technologies are overall in the early stages of development, and companies need to be patient, understanding the experiments and proof of concepts will need to be updated in the future.
\item	Once companies are fully aware of the value and impact of the technology, a strategic roadmap will be developed, critical to making the business quantum-ready.
\item	The mastering of the technology will come with a full implementation within the business.
\end{enumerate}

End users are the ultimate beneficiaries of quantum technologies and their engagement is integral to shaping practical applications and driving market demand. They can play an important role in defining the use cases for quantum technologies.

Industries contribute to the quantum ecosystem by providing feedback and helping to fine-tune quantum technologies. As early adopters, they can trial quantum solutions, thereby identifying gaps, suggesting improvements, and ultimately aiding in the refinement and evolution of these technologies. Their feedback forms an iterative loop with developers, accelerating the maturity of quantum solutions. By participating in industry forums and public consultations, end users can help shape the discourse around regulatory standards, ethical guidelines, and future research priorities. It's also worth noting that engagement with industries needs to be a two-way street. Technologists and researchers need to understand the needs and constraints of end users, while end users need to be educated about the potential and limitations of quantum technologies. This mutual understanding can help to ensure that quantum technologies are developed in a way that is both innovative and practically beneficial. All in all, becoming quantum-ready for a business means taking full advantage of the technology.

\subsection{Academia and skill-based institutions}

Being quantum-ready involves a combination of both theoretical knowledge and practical skills needed to work in the rapidly evolving field of quantum technologies. 

Academic institutions can play an important role in filling the gaps needed to be quantum-ready by providing individuals with the foundational knowledge and transferable skills through specialised academic programs, conducting research, partnering with industry for hands-on training, providing experiential learning opportunities, and offering continuing education programs. 

While it may be challenging for academic programs to have enough resources to make students fully qualified for quantum careers, many programs are taking steps to address this challenge and ensure that their students are well-prepared for the demands of the field. For example, some universities and colleges are establishing dedicated quantum computing centres, institutes, and labs to provide students with access to state-of-the-art equipment and resources \cite{qureca_MS, 9705217, Gabriele2021}. These centres often work closely with industry partners to ensure that students are gaining the skills and knowledge that are most in demand in the field.

An increasing number of academic stakeholders and associations worldwide are starting to create and include specialised courses in quantum technologies within the syllabus of interdisciplinary programs and degrees.

To create a diverse quantum workforce and mitigate potential biases in quantum technologies, academic programs need to attract individuals from underrepresented groups to fields such as computer science, math, statistics, physics, material science, and chemistry. These efforts should begin early, with K-12 interventions that offer exposure to quantum technologies and make them accessible to all students. Partnering with online learning platforms and skill-based training institutions can be another way for academic institutions to provide students with access to a broader range of resources and expertise \cite{qureca,qusteam}. Programs like \emph{Qubit by Qubit} have seen success in increasing interest and participation in quantum computing among middle and high school students from diverse backgrounds \cite{qbq}. Universities also have a crucial role in building a talent pipeline by offering courses, programs, and research opportunities that provide students with the necessary skills and knowledge to pursue quantum careers. In addition, business leaders can influence the shift towards quantum learning in schools and universities by demonstrating the importance of this field through corporate involvement. Ultimately, academic programs must work to create a pathway for the talent pipeline by removing barriers to entry and ensuring that all students have access to resources and opportunities to develop quantum-ready skills \cite{Kaur}. 

The field of quantum technologies requires a diverse workforce to meet its needs. However, STEM disciplines still face challenges in broadening participation, particularly among women and underrepresented racial and ethnic groups. Academic programs must address the barriers that currently exist and provide support, mentorship, and recognition for diverse perspectives and expertise, to increase accessibility to students and to promote inclusivity and diversity \cite{Aiello_2021}.

\subsection{Readiness with respect to ethics and protocols}

Until this section, a general framework for quantum readiness and the elaboration for different stakeholders have been provided. Being quantum-ready involves several factors, including the availability of quantum hardware and software, the expertise of personnel, and the quality of the infrastructure. To be quantum-ready, an organisation must have access to the necessary technology, personnel, and infrastructure. However, this narrow concept of quantum readiness can, and sometimes should, be expanded to cover any ethical and legal aspects, especially following the national and international protocols, such as those on standardisation.

Being quantum-ready is an important step for organisations that are considering the use of quantum technologies. It allows organisations to identify and address any challenges or limitations that may prevent them from effectively using quantum technologies. For example, an organisation that is not quantum-ready may lack the necessary hardware, software, expertise, or infrastructure to effectively use quantum technologies. In order to be quantum-ready, an organisation must invest in quantum technologies, by developing new algorithms and/or applications, training personnel, and building a quantum-safe infrastructure. 

Investing in quantum readiness is a significant undertaking and requires a long-term commitment to preparing for the use of quantum technologies. It is not something that can be done quickly or easily. However, the benefits of being quantum-ready are significant. Organisations that are quantum-ready will be able to take advantage of the potential benefits of quantum technologies and will be well-positioned to compete in the emerging quantum global economy.

Companies, governments, NGOs, and other stakeholders must be quantum-ready because quantum technologies promise some essentially different capabilities. To take advantage of the benefits of quantum technologies, organisations must be prepared to use them effectively. However, quantum technologies raise many ethical concerns, including the potential misuse of technology, its impact on jobs and the economy, unequal access to technology, and its potential impact on society. 

As one well-known example, quantum technologies could be used for malicious purposes such as hacking and espionage, which highlights the need for appropriate safeguards and regulations \cite{mosca2022}. This directly impacts the meaning of quantum readiness. Additionally, the use of quantum computing could automatise specific tasks, potentially leading to job losses in certain industries. This raises concerns about the potential for economic disruption and the need for policies to support affected workers and sectors. One can argue whether being quantum-ready also encompasses being ready against such disruptions brought forth by this new and emerging set of technologies. Furthermore, only certain individuals or organisations may have access to mature quantum technologies \cite{wef22}, leading to potentially increasing inequalities and disparities. This raises concerns about fairness and equitable access to the benefits of quantum technologies and the need for policies to ensure that the technology is accessible to all. Finally, there are also concerns about the potential impact of quantum technologies on different stakeholder groups in society \cite{deWolf2017}, and the need for careful and deliberate planning to ensure that quantum technologies can be used ethically and beneficially.

There are many discussions and a growing literature on ethical and societal responsible use of quantum technologies ranging from arguments on the democratisation of quantum technologies \cite{seskir2023} to calls for not even building quantum computers at all \cite{mckay2022}. In practical terms, these discussions inform organisations, governments and the public, and could change the stakeholders' perception of quantum technologies. For example, the topic of trust is discussed openly in a Time article \cite{campbell2023}, where it was proposed that "...soon will come a time when trusting a quantum computer will require a leap of faith", and without trust-building across the entire ecosystem, there will be strong hesitancy to take this leap of faith. There are efforts to mitigate this hesitancy and widely accepted ethical guidelines and protocols are established methods of trust-building in new and emerging technology ecosystems.

To employ methods of trust-building, one needs to frame the object of this trust, in this case, quantum technologies. It is discussed in the literature that quantum technologies can be considered system technologies with a systemic impact on society \cite{deJong2022}, which calls for an anticipatory approach. Similarly, arguing for a strong \emph{Responsible Research and Innovation} (RRI) approach can also be found in the literature \cite{coenen2017}. There is even a call for action from some industry partners and different stakeholders in the ecosystem for the ethical development of these technologies \cite{TQI2021}. There are also many proposals and important highlights in the literature on arguing for further understandability of the quantum theory \cite{vermaas2017}, the effects of hype on teaching about quantum technologies \cite{meyer2023}, introducing ethics into quantum information classrooms \cite{meyer2022}, awareness raising as the bare minimum that should be done \cite{roberson2021}, and many other potential actions to prevent quantum technologies running into fiascos of implementation at the interface of science and society \cite{coenen2022}.
One might consider how these discussions can be practically tied to an organisation's quantum readiness. A clear example of this is the White House memorandum \textit{on Promoting United States Leadership in Quantum Computing While Mitigating Risk to Vulnerable Cryptographic Systems}, which prioritises "...the timely and equitable transition of cryptographic systems to quantum-resistant cryptography, with the goal of mitigating as much of the quantum risk as is feasible by 2035"
\cite{memorandum2022}. For those who have been following the discussions on the quantum threat \cite{mosca2022}, this came as no surprise. It is the poster child case of the potential of future quantum technologies causing a massive impact on today's regulatory landscape. It might seem like this was always supposed to happen, but the chances of this becoming a regulation, let alone a company fundraising \$500 million on this particular topic was negligible just a decade ago \cite{sandbox500}. Furthermore, this is a particular example of how the values and priorities of certain stakeholders in the ecosystem can cause major shifts in not only the narratives but also the regulatory and opportunity landscape of an entire industry with consequences far-reaching into almost all industries.

There are many moving parts on the regulatory landscape both in early and late stage standardisation efforts such as the Quantum Internet Research Group \cite{qirg} under the Internet Engineering Task Force working on the future standards of the quantum internet while others such as \emph{European Telecommunications Standards Institute} (ETSI) \emph{Industry Specification Group} (ISG) working on QKD standards \cite{etsiqkd}, which has an impact on industrial relations much sooner. Ethics plays an important role in these processes. Different stakeholders have different value sets and even within the stakeholder groups with similar value sets, there are differences in prioritisation of these. There are some promoting global open access and sustainability while not explicitly for democratic values due to geopolitical concerns, while others that argue strongly for democratisation may find themselves only arguing for it in a limited context of like-minded countries. Discussions on export controls and even controls on knowledge transfer are common practices in the ethical and regulatory context of quantum technologies. 

As discussed earlier in this article, techno-economic paradigms are characterised by a set of shared beliefs, values, and practices that define the boundaries of what is considered technically feasible and economically viable \cite{dosi1998}, but in a wider context, these also need to be societal acceptable \cite{perez2010} to fully become system technologies \cite{deJong2022}. Hence, organisations aiming for quantum readiness should keep track and maybe even actively participate in these discussions on what should be the wider set of values and protocols surrounding how these technologies interface with the market and the whole of society \cite{coenen2022}.

\subsection{People and Society}
People and society play pivotal roles in shaping and sustaining the quantum ecosystem \cite{canada}. Beyond education, the public dialogue and consensus-building processes are crucial to address the ethical, legal, and societal implications of quantum technologies. Quantum cryptography's effects on data privacy, the implications of quantum computing on job security and new forms of employment, and other such considerations should be widely and openly discussed. From a societal perspective, fostering an environment that encourages innovation and entrepreneurship in quantum technologies is crucial. Policymakers, guided by public sentiment and expert advice, have a critical role in shaping regulations that will determine how quantum technologies are used, ensuring their alignment with societal values and norms. As we explore the role of end users in promoting and normalising the adoption of quantum technologies, it becomes evident that their experiences and testimonials can be of considerable value in inspiring confidence and interest in other potential users and significantly influence the widespread adoption of quantum technologies through their advocacy and normalisation efforts.

Nurturing a robust and effective quantum ecosystem is a collective endeavor that demands the synchronised efforts of diverse stakeholders. As quantum technology transcends the borders of academia and finds its footing in various industries, it is incumbent upon all participants to fulfill their roles conscientiously. Researchers, educators, industry practitioners, policymakers, and investors all bear the mantle of responsibility to shape a quantum-ready future. As we stand on the cusp of this second quantum revolution, it becomes crucial to leverage the power of this integrated ecosystem to transform the theoretical potential of quantum technology into tangible benefits for society at large.

\section{Conclusion}

Quantum readiness refers to the state of being equipped to adopt and utilise quantum technology, providing organisations and nations with the potential to secure a first-mover advantage, shape the technology's future direction, attain a competitive edge, and achieve economic prosperity. This paper highlights the importance of quantum readiness, along with innovation models and assessment tools that enable individuals, organisations, and businesses to evaluate their quantum readiness and formulate strategies for becoming quantum-ready.

We started with the classification of emerging technologies based on their level of maturity and potential impact which provides a useful framework for policymakers and business leaders to understand and prepare for the impact of these technologies on their respective industries. Disruptive technologies have the potential to create entirely new markets or disrupt existing ones while enabling technologies facilitate the development of new goods, services, or business models. Sustaining technologies, on the other hand, improve the performance of existing products or services without necessarily creating new markets or disrupting existing ones. Quantum technology is an emerging disruptive technology that represents a significant departure from traditional classical technologies and has the potential to revolutionise various industries. Policymakers and business leaders can position themselves to benefit from the potential of quantum technology by proactively identifying potential disruptions, determining areas for research and development, and informing policy decisions. 

In the third section, we addressed the need to understand quantum readiness. It is anticipated that the rapidly advancing quantum technology will disrupt various industry sectors, revolutionise cryptography, improve cybersecurity, and accelerate scientific research. Countries and organisations that invest in and embrace quantum technology will have a significant competitive advantage, while those that do not may fall behind. Quantum readiness is crucial for businesses and governments to thrive in the future, and addressing this issue requires time, research and development investments, skilled human resources, and strategic decisions. Quantum technology shares characteristics with \emph{General-Purpose Technologies} (GPTs), which have a long-term impact on the economy and society. Therefore, organisations can ensure their competitiveness and economic prosperity by collaborating to build a strong quantum ecosystem and embracing quantum technology.

We formalised the \emph{Quantum Technology Readiness Levels} (QTRLs), their descriptions, and expected outcomes on the basis of which we assessed the current status of different quantum technologies. We also established a time horizon indicating the anticipated date by which these technologies will be successfully adopted and deployed for practical applications. Unsurprisingly, quantum computing is still in its infancy; by contrast, quantum sensing and communications can be considered to be in more advanced stages.

We then introduced Quantum Commercial Readiness Levels (QCRLs) as a complementary framework to Quantum Technology Readiness Levels (QTRLs). By doing so, we not only address the technical maturity of quantum technologies but also their market readiness and commercial viability. Furthermore, we have explored the process of mapping QCRLs to QTRLs. This brings a necessary cohesion between the technical progression and market readiness of quantum technology.

Finally, we discussed the significance of quantum readiness for key ecosystem stakeholders and defined relevant indicators to be quantum-ready.  

\begin{enumerate}
    \item Governments need to establish infrastructure, allocate resources, and develop expertise to become quantum-ready. National quantum strategies and roadmaps can assist in identifying priority innovation areas for investment. An effective local ecosystem is also important, requiring partnerships with businesses, research institutions, and supporting organisations to ensure a secure and robust supply chain. To address ethical and policy challenges, governments must implement standards and certifications to ensure the security and reliability of quantum technologies.
    \item The process of becoming quantum-ready for a company depends on its size and its willingness to embrace new technologies. The approach involves first understanding the basics of quantum technology and getting involved in the ecosystem by utilising available resources. At a later stage, identifying use cases and experimenting with proof of concept with quantum startups brings a competitive advantage to the business. Once the company comprehends the benefits of the technology, a strategic roadmap is established for complete integration within the business. Proficiency in the technology ultimately enables the company to harness its full potential.
    \item The development of a quantum workforce with foundational knowledge and transferable skills required for the rapidly evolving field of quantum technologies can be facilitated by academic and skill-based institutions through specialised academic programs, research, hands-on training, experiential learning opportunities, and continuing education programs. To promote inclusivity and diversity, academic programs must target individuals from underrepresented groups in fields such as computer science, math, statistics, physics, material science, and chemistry, beginning with K-12 interventions. Additionally, academic institutions need to establish a pathway for the talent pipeline by removing barriers to entry and providing resources and opportunities for students to acquire quantum-ready skills. To increase accessibility and inclusivity, academic programs must address the barriers that currently exist and provide support, mentorship, and recognition for diverse perspectives and expertise.
    \item The benefits of being quantum-ready are significant, allowing organisations to take advantage of the potential benefits of quantum technologies and be well-positioned to compete in the emerging quantum global economy. However, there are also ethical and legal aspects to consider, and organisations must follow national and international protocols, such as those on standardisation. Quantum technologies may also raise ethical concerns, including potential misuse, impact on jobs and the economy, unequal access to technology, and its impact on society. There is growing attention towards the ethical and societal responsible use of quantum technologies, ranging from arguments on the democratisation of quantum technologies to calls for not building quantum computers at all. To employ methods of trust-building, organisations need to frame the object of this trust, in this case, quantum technologies, and take an anticipatory approach. Furthermore, there is a growing need for a call for action from industry partners and different stakeholders in the ecosystem to take into account the ethical and societal concerns to effectively use quantum technologies.
\end{enumerate}

In the coming years, the quantum technology sector is expected to undergo significant changes. Some technologies will become operational, while others may not succeed due to challenges in engineering or commercialisation. It is not possible to make precise predictions about which technologies will succeed or fail. However, adopting a forward-looking approach can help organisations or individuals become quantum-ready.

\insertbibliography{References}


\begin{thebibliography}{10}
\providecommand{\url}[1]{#1}
\csname url@samestyle\endcsname
\providecommand{\newblock}{\relax}
\providecommand{\bibinfo}[2]{#2}
\providecommand{\BIBentrySTDinterwordspacing}{\spaceskip=0pt\relax}
\providecommand{\BIBentryALTinterwordstretchfactor}{4}
\providecommand{\BIBentryALTinterwordspacing}{\spaceskip=\fontdimen2\font plus
\BIBentryALTinterwordstretchfactor\fontdimen3\font minus
  \fontdimen4\font\relax}
\providecommand{\BIBforeignlanguage}[2]{{%
\expandafter\ifx\csname l@#1\endcsname\relax
\typeout{** WARNING: IEEEtran.bst: No hyphenation pattern has been}%
\typeout{** loaded for the language `#1'. Using the pattern for}%
\typeout{** the default language instead.}%
\else
\language=\csname l@#1\endcsname
\fi
#2}}
\providecommand{\BIBdecl}{\relax}
\BIBdecl

\bibitem{Seskir2022}
\BIBentryALTinterwordspacing
Z.~C. Seskir, R.~Korkmaz, and A.~U. Aydinoglu, ``The landscape of the quantum
  start-up ecosystem,'' \emph{EPJ Quantum Technology}, vol.~9, no.~1, p.~27,
  Oct 2022. [Online]. Available:
  \url{https://doi.org/10.1140/epjqt/s40507-022-00146-x}
\BIBentrySTDinterwordspacing

\bibitem{qureca2023}
M.~Kaur. (2023) Overview of quantum initiatives worldwide 2023.
  \url{https://qureca.com/overview-of-quantum-initiatives-worldwide-2023/}
  (Retrieved on: 19/06/2023).

\bibitem{ey}
H.~Lewis. (2022) How can you prepare now for the quantum computing future?
  \url{https://www.ey.com/en_uk/emerging-technologies/how-can-you-prepare-now-for-the-quantum-computing-future}.

\bibitem{bell}
E.~Campbell, B.~Terhal, and C.~Vuillot, ``Roads towards fault-tolerant
  universal quantum computation,'' \emph{Nature}, vol. 549, pp. 172--179, 09
  2017.

\bibitem{chia}
J.~Chiaverini, D.~Leibfried, T.~Schätz, M.~Barrett, R.~Blakestad, J.~Britton,
  W.~Itano, J.~Jost, E.~Knill, C.~Langer, R.~Ozeri, and D.~Wineland,
  ``Realization of quantum error correction,'' \emph{Nature}, vol. 432, pp.
  602--5, 01 2005.

\bibitem{seb}
F.~Sebastiano, H.~Homulle, B.~Patra, R.~Incandela, J.~Dijk, L.~Song, M.~Babaie,
  A.~Vladimirescu, and E.~Charbon, ``Cryo-cmos electronic control for scalable
  quantum computing: Invited,'' 06 2017, pp. 1--6.

\bibitem{antonio}
L.~Sapienza, M.~Davanco, A.~Badolato, and K.~Srinivasan, ``Optical positioning
  of single-photon emitters for quantum information technology applications,''
  01 2017, p. QW3C.2.

\bibitem{anton}
A.~Badolato, ``Nanocavities, artificial atoms, and photons: Quantum optics in
  the nanoworld,'' 01 2014, p. LTu1B.1.

\bibitem{rotolo}
D.~Rotolo, D.~Hicks, and B.~Martin, ``What is an emerging technology?''
  \emph{SSRN Electronic Journal}, 01 2015.

\bibitem{freeman}
\BIBentryALTinterwordspacing
F.~L. Chris~Freeman, ``As time goes by: From the industrial revolutions to the
  information revolution,'' \emph{Oxford University Press}, p. 424, 2001.
  [Online]. Available:
  \url{https://global.oup.com/academic/product/as-time-goes-by-9780199241071?cc=gb&lang=en&#}
\BIBentrySTDinterwordspacing

\bibitem{miles}
I.~Miles and K.~Fursov, ``Framing emerging nanotechnologies: Sketches towards a
  forward-looking analysis of skills,'' \emph{Higher School of Economics, Basic
  Research Program Working Papers - Series: Science, Technology And
  Innovation}, vol. WP BRP 15/STI/2013, 07 2013.

\bibitem{man}
C.~Freeman and L.~Soete, \emph{The Economics of Industrial Innovation}, 01
  1997, vol.~1.

\bibitem{oro}
T.~Oroszi, ``Disruption innovation and theory,'' \emph{Journal of Service
  Science and Management}, vol.~13, pp. 449--458, 01 2020.

\bibitem{kumars}
A.~Kumaraswamy, R.~Garud, and S.~Ansari, ``Perspectives on disruptive
  innovations,'' \emph{SSRN Electronic Journal}, 01 2018.

\bibitem{coccia}
M.~Coccia, ``Disruptive innovations in quantum technologies,'' vol.~9, pp.
  21--39, 05 2022.

\bibitem{sch}
T.~Scheidsteger, R.~Haunschild, L.~Bornmann, and C.~Ettl, ``Quantum technology
  2.0 -- topics and contributing countries from 1980 to 2018,'' 05 2021.

\bibitem{shelli}
\BIBentryALTinterwordspacing
S.~Brunswick, ``Quantum technology and the global space ecosystem,''
  \emph{Forbes}, 2022. [Online]. Available:
  \url{https://www.forbes.com/sites/forbestechcouncil/2022/12/15/quantum-technology-and-the-global-space-ecosystem/?sh=77cc0fc44a68}
\BIBentrySTDinterwordspacing

\bibitem{dow}
J.~Dowling and G.~Milburn, ``Quantum technology: The second quantum
  revolution,'' \emph{Philosophical transactions. Series A, Mathematical,
  physical, and engineering sciences}, vol. 361, pp. 1655--74, 09 2003.

\bibitem{perrier}
E.~Perrier, ``The quantum governance stack: Models of governance for quantum
  information technologies,'' \emph{Digital Society}, vol.~1, 10 2022.

\bibitem{sidhu}
\BIBentryALTinterwordspacing
J.~S. Sidhu, S.~K. Joshi, M.~Gündoğan, T.~Brougham, D.~Lowndes,
  L.~Mazzarella, M.~Krutzik, S.~Mohapatra, D.~Dequal, G.~Vallone, P.~Villoresi,
  A.~Ling, T.~Jennewein, M.~Mohageg, J.~G. Rarity, I.~Fuentes, S.~Pirandola,
  and D.~K.~L. Oi, ``Advances in space quantum communications,'' \emph{IET
  Quantum Communication}, vol.~2, no.~4, pp. 182--217, 2021. [Online].
  Available:
  \url{https://ietresearch.onlinelibrary.wiley.com/doi/abs/10.1049/qtc2.12015}
\BIBentrySTDinterwordspacing

\bibitem{kilber}
N.~Kilber, D.~Kaestle, and S.~Wagner, ``Cybersecurity for quantum computing,''
  10 2021.

\bibitem{qrt}
``Quantum readiness toolkit: Building a quantum-secure economy | world economic
  forum,''
  \url{https://www.weforum.org/whitepapers/quantum-readiness-toolkit-building-a-quantum-secure-economy}.

\bibitem{zinner}
M.~Zinner, F.~Dahlhausen, P.~Boehme, J.~P. Ehlers, L.~Bieske, and L.~Fehring,
  ``Quantum computing's potential for drug discovery: Early stage industry
  dynamics,'' \emph{Drug Discovery Today}, vol.~26, 06 2021.

\bibitem{krenn}
M.~Krenn, J.~Landgraf, T.~Foesel, and F.~Marquardt, ``Artificial intelligence
  and machine learning for quantum technologies,'' 08 2022.

\bibitem{bauer}
B.~Bauer, S.~Bravyi, M.~Motta, and G.~Chan, ``Quantum algorithms for quantum
  chemistry and quantum materials science,'' 10 2020.

\bibitem{egger}
D.~Egger, C.~Gambella, J.~Marecek, S.~McFaddin, M.~Mevissen, R.~Raymond,
  A.~Simonetto, S.~Woerner, and E.~Yndurain, ``Quantum computing for finance:
  State-of-the-art and future prospects,'' \emph{IEEE Transactions on Quantum
  Engineering}, vol.~1, 10 2020.

\bibitem{berger}
C.~Berger, A.~Paolo, T.~Forrest, S.~Hadfield, N.~Sawaya, M.~Stęchły, and
  K.~Thibault, ``Quantum technologies for climate change: Preliminary
  assessment,'' 06 2021.

\bibitem{diet}
J.~Dieterich and E.~Carter, ``Opinion: Quantum solutions for a sustainable
  energy future,'' \emph{Nature Reviews Chemistry}, vol.~1, p. 0032, 04 2017.

\bibitem{parrish}
C.~Parrish~II, D.~Hebert, A.~Jackson, K.~Ramasamy, H.~McDaniel, G.~Giacomelli,
  and M.~Bergren, ``Optimizing spectral quality with quantum dots to enhance
  crop yield in controlled environments,'' \emph{Communications Biology},
  vol.~4, p. 124, 01 2021.

\bibitem{cid}
M.~Cid, J.~González, L.~Martín, and D.~Gómez, ``Disruptive quantum safe
  technologies,'' 08 2022, pp. 1--8.

\bibitem{deo}
J.~Deodoro, M.~Gorbanyov, M.~Malaika, and T.~Saadi~Sedik, ``Quantum computing
  and the financial system: Spooky action at a distance?'' 03 2021.

\bibitem{cao}
Y.~Cao, J.~Fontalvo, and A.~Aspuru-Guzik, ``Potential of quantum computing for
  drug discovery,'' \emph{IBM Journal of Research and Development}, vol.~PP,
  pp. 1--1, 12 2018.

\bibitem{Christensen1997}
C.~M. Christensen, \emph{{The Innovator's Dilemma: When New Technologies Cause
  Great Firms to Fail}}.\hskip 1em plus 0.5em minus 0.4em\relax {Harvard
  Business Review Press}, 1997.

\bibitem{markides2013}
C.~C. Markides, ``Business model innovation: What can the ambidexterity
  literature teach us?'' \emph{Academy of Management Perspectives}, vol.~27,
  no.~4, p. 313–323, 2013.

\bibitem{perez2010}
\BIBentryALTinterwordspacing
C.~Perez, ``{Technological revolutions and techno-economic paradigms},''
  \emph{Cambridge Journal of Economics}, vol.~34, no.~1, pp. 185--202, 09 2009.
  [Online]. Available: \url{https://doi.org/10.1093/cje/bep051}
\BIBentrySTDinterwordspacing

\bibitem{dosi1998}
G.~Dosi, C.~Freeman, R.~Nelson, G.~Silverberg, and L.~Soete, \emph{Technical
  change and economic theory}.\hskip 1em plus 0.5em minus 0.4em\relax Pinter
  Publ, 1988.

\bibitem{geels2017}
F.~W. Geels, B.~K. Sovacool, T.~Schwanen, and S.~Sorrell, ``Sociotechnical
  transitions for deep decarbonization,'' \emph{Science}, vol. 357, no. 6357,
  p. 1242–1244, 2017.

\bibitem{murat}
\BIBentryALTinterwordspacing
M.~Bengisu and R.~Nekhili, ``Forecasting emerging technologies with the aid of
  science and technology databases,'' \emph{Technological Forecasting and
  Social Change}, vol.~73, no.~7, pp. 835--844, 2006. [Online]. Available:
  \url{https://www.sciencedirect.com/science/article/pii/S0040162505001393}
\BIBentrySTDinterwordspacing

\bibitem{mankins}
J.~Mankins, ``Technology readiness level – a white paper,'' 01 1995.

\bibitem{conrow}
E.~Conrow, ``Estimating technology readiness level coefficients,''
  \emph{Journal of Spacecraft and Rockets - J SPACECRAFT ROCKET}, vol.~48, pp.
  146--152, 01 2011.

\bibitem{ani}
I.~Animah and M.~Shafiee, ``A framework for assessment of technological
  readiness level (trl) and commercial readiness index (cri) of asset
  end-of-life strategies,'' 06 2018.

\bibitem{sawag}
\BIBentryALTinterwordspacing
M.~Sawaguchi, ``Innovation activities based on s-curve analysis and patterns of
  technical evolution-“from the standpoint of engineers, what is
  innovation?”,'' \emph{Procedia Engineering}, vol.~9, pp. 596--610, 2011,
  proceeding of the ETRIA World TRIZ Future Conference. [Online]. Available:
  \url{https://www.sciencedirect.com/science/article/pii/S1877705811001627}
\BIBentrySTDinterwordspacing

\bibitem{maria}
\BIBentryALTinterwordspacing
T.~J. S. . T.~T. Maria~Priestley, ``Innovation on the web: the end of the
  s-curve?'' \emph{Internet Histories}, 2020. [Online]. Available:
  \url{https://eprints.soton.ac.uk/439144/1/paper_published.pdf}
\BIBentrySTDinterwordspacing

\bibitem{li}
Y.~Li and S.~Qin, ``Comparative study on the developmental stages of global ccs
  technology based on the s-curve model,'' \emph{Energy RESEARCH LETTERS},
  vol.~2, 08 2021.

\bibitem{vernon}
\BIBentryALTinterwordspacing
R.~Vernon, ``{International Investment and International Trade in the Product
  Cycle*},'' \emph{The Quarterly Journal of Economics}, vol.~80, no.~2, pp.
  190--207, 05 1966. [Online]. Available: \url{https://doi.org/10.2307/1880689}
\BIBentrySTDinterwordspacing

\bibitem{cawson}
A.~Cawson, L.~Haddon, and I.~Miles, \emph{The Shape of Things to Consume:
  Delivering Information Technology into the Home}, 01 1995.

\bibitem{utb}
J.~M. Utterback, \emph{Mastering the dynamics of innovation: how companies can
  seize opportunities in the face of technological change}, 253 1994.

\bibitem{rueda}
G.~Rueda and D.~F. Kocaoglu, ``Diffusion of emerging technologies: An
  innovative mixing approach,'' in \emph{PICMET '08 - 2008 Portland
  International Conference on Management of Engineering \& Technology}, 2008,
  pp. 672--697.

\bibitem{Seskir2021}
\BIBentryALTinterwordspacing
Z.~C. Seskir and A.~U. Aydinoglu, ``The landscape of academic literature in
  quantum technologies,'' \emph{International Journal of Quantum Information},
  vol.~19, no.~02, p. 2150012, 2021. [Online]. Available:
  \url{https://doi.org/10.1142/S021974992150012X}
\BIBentrySTDinterwordspacing

\bibitem{seskir_patent_2023}
Z.~C. Seskir and K.~W. Willoughby, ``Global innovation and competition in
  quantum technology, viewed through the lens of patents and artificial
  intelligence,'' \emph{International Journal of Intellectual Property
  Management}, vol.~13, no.~1, p.~40, 2023.

\bibitem{matt}
\BIBentryALTinterwordspacing
M.~Swayne, ``Quantum technology 2022 investment update - key trends and
  players,'' \emph{The Quantum Insider}, 2022. [Online]. Available:
  \url{https://thequantuminsider.com/2023/02/17/quantum-technology-2022-investment-update-key-trends-and-players/}
\BIBentrySTDinterwordspacing

\bibitem{ole}
A.~Olechowski, S.~Eppinger, N.~Joglekar, and K.~Tomaschek, ``Technology
  readiness levels: Shortcomings and improvement opportunities,'' \emph{Systems
  Engineering}, vol.~23, 03 2020.

\bibitem{h2020}
\BIBentryALTinterwordspacing
E.~C.~D. C, ``Horizon 2020 work programme,'' \emph{European commission}, p.~39,
  2017. [Online]. Available:
  \url{https://ec.europa.eu/research/participants/data/ref/h2020/other/wp/2016-2017/annexes/h2020-wp1617-annex-ga_en.pdf}
\BIBentrySTDinterwordspacing

\bibitem{trl0}
UKRI. (2022) Eligibility of technology readiness levels (trl).
  \url{https://www.ukri.org/councils/stfc/guidance-for-applicants/check-if-youre-eligible-for-funding/eligibility-of-technology-readiness-levels-trl/}.

\bibitem{jose}
\BIBentryALTinterwordspacing
G.~G.~E. Martinez Plumed~Fernado and H.-O. Jose, ``Ai watch: Assessing
  technology readiness levels for artificial intelligence,'' \emph{JRC
  Publications Repository}, 2020. [Online]. Available:
  \url{https://publications.jrc.ec.europa.eu/repository/handle/JRC122014}
\BIBentrySTDinterwordspacing

\bibitem{fzj}
F.~Jülich. (2022) Technology readiness level of quantum computing technology
  (qtrl).
  \url{https://www.fz-juelich.de/en/ias/jsc/about-us/structure/research-groups/qip/technology-readiness-level-of-quantum-computing-technology-qtrl?expand=translations,fzjsettings,nearest-institut}.

\bibitem{andi}
A.~Sama. (2022) Quantum.tech 2022: Next generation insights of technology
  innovations.

\bibitem{krelina}
M.~Krelina, ``Quantum technology for military applications,'' \emph{EPJ Quantum
  Technology}, vol.~8, 12 2021.

\bibitem{greig}
C.~Greig, G.~Bongers, C.~Stott, and S.~Byrom, ``Overview of ccs roadmaps and
  projects,'' 01 2016.

\bibitem{nemet}
G.~Nemet, V.~Zipperer, and M.~Kraus, ``The valley of death, the technology pork
  barrel, and public support for large demonstration projects,'' \emph{Energy
  Policy}, vol. 119, pp. 154--167, 08 2018.

\bibitem{Kaur}
\BIBentryALTinterwordspacing
M.~Kaur and A.~Venegas-Gomez, ``{Defining the quantum workforce landscape: a
  review of global quantum education initiatives},'' \emph{Optical
  Engineering}, vol.~61, no.~8, p. 081806, 2022. [Online]. Available:
  \url{https://doi.org/10.1117/1.OE.61.8.081806}
\BIBentrySTDinterwordspacing

\bibitem{sodhi}
\BIBentryALTinterwordspacing
M.~Sodhi and S.~Tayur, ``Make your business quantum-ready today,''
  \emph{Management and Business Review}, April 2022, copyright {\copyright}
  2022 Management and Business Review. All Rights Reserved. [Online].
  Available: \url{https://openaccess.city.ac.uk/id/eprint/28064/}
\BIBentrySTDinterwordspacing

\bibitem{qureca_MS}
QURECA. (2019) Masters in quantum technologies.
  \url{https://qureca.com/masters-in-quantum-technologies/} (Retrieved on:
  27/02/2023).

\bibitem{9705217}
\BIBentryALTinterwordspacing
A.~Asfaw, A.~Blais, K.~R. Brown, J.~Candelaria, C.~Cantwell, L.~D. Carr,
  J.~Combes, D.~M. Debroy, J.~M. Donohue, S.~E. Economou, E.~Edwards, M.~F.~J.
  Fox, S.~M. Girvin, A.~Ho, H.~M. Hurst, Z.~Jacob, B.~R. Johnson,
  E.~Johnston-Halperin, R.~Joynt, E.~Kapit, J.~Klein-Seetharaman, M.~Laforest,
  H.~J. Lewandowski, T.~W. Lynn, C.~R.~H. McRae, C.~Merzbacher, S.~Michalakis,
  P.~Narang, W.~D. Oliver, J.~Palsberg, D.~P. Pappas, M.~G. Raymer, D.~J.
  Reilly, M.~Saffman, T.~A. Searles, J.~H. Shapiro, and C.~Singh, ``Building a
  quantum engineering undergraduate program,'' \emph{IEEE Transactions on
  Education}, vol.~65, no.~2, pp. 220--242, 2022. [Online]. Available:
  \url{https://doi.org/10.1109/TE.2022.3144943}
\BIBentrySTDinterwordspacing

\bibitem{Gabriele2021}
\BIBentryALTinterwordspacing
G.~Rainò, L.~Novotny, and M.~Frimmer, ``Quantum engineers in high demand,''
  \emph{Nature Materials}, vol.~20, p. 1449, 2021. [Online]. Available:
  \url{https://doi.org/10.1038/s41563-021-01080-6}
\BIBentrySTDinterwordspacing

\bibitem{qureca}
QURECA. (2019) Qureca (quantum resources \& careers) training.
  \url{https://qureca.com/quantum-careeers-training/} (Retrieved on:
  27/02/2023).

\bibitem{qusteam}
QUSTEAM. (2021) Qusteam initiative. \url{https://qusteam.org/} (Retrieved on:
  27/02/2023).

\bibitem{qbq}
T.~C. School. (2014) Qubit by qubit. \url{https://www.qubitbyqubit.org/}
  (Retrieved on: 27/02/2023).

\bibitem{Aiello_2021}
\BIBentryALTinterwordspacing
C.~D. Aiello, D.~D. Awschalom, H.~Bernien, T.~Brower, K.~R. Brown, T.~A. Brun,
  J.~R. Caram, E.~Chitambar, R.~D. Felice, K.~M. Edmonds, M.~F.~J. Fox,
  S.~Haas, A.~W. Holleitner, E.~R. Hudson, J.~H. Hunt, R.~Joynt, S.~Koziol,
  M.~Larsen, H.~J. Lewandowski, D.~T. McClure, J.~Palsberg, G.~Passante, K.~L.
  Pudenz, C.~J.~K. Richardson, J.~L. Rosenberg, R.~S. Ross, M.~Saffman,
  M.~Singh, D.~W. Steuerman, C.~Stark, J.~Thijssen, A.~N. Vamivakas, J.~D.
  Whitfield, and B.~M. Zwickl, ``Achieving a quantum smart workforce,''
  \emph{Quantum Science and Technology}, vol.~6, no.~3, p. 030501, jun 2021.
  [Online]. Available: \url{https://dx.doi.org/10.1088/2058-9565/abfa64}
\BIBentrySTDinterwordspacing

\bibitem{mosca2022}
M.~Mosca and M.~Piani. (2022) Quantum threat timeline report 2022.
  \url{https://globalriskinstitute.org/publication/2022-quantum-threat-timeline-report/}
  (Retrieved on: 27/02/2023).

\bibitem{wef22}
\BIBentryALTinterwordspacing
``World economic forum releases quantum computing guidelines,'' \emph{MRS
  Bulletin}, vol.~47, no.~4, pp. 355--356, Apr 2022. [Online]. Available:
  \url{https://doi.org/10.1557/s43577-022-00331-4}
\BIBentrySTDinterwordspacing

\bibitem{deWolf2017}
\BIBentryALTinterwordspacing
R.~de~Wolf, ``The potential impact of quantum computers on society,''
  \emph{Ethics and Information Technology}, vol.~19, no.~4, pp. 271--276, Dec
  2017. [Online]. Available: \url{https://doi.org/10.1007/s10676-017-9439-z}
\BIBentrySTDinterwordspacing

\bibitem{seskir2023}
\BIBentryALTinterwordspacing
Z.~C. Seskir, S.~Umbrello, C.~Coenen, and P.~E. Vermaas, ``Democratization of
  quantum technologies,'' \emph{Quantum Science and Technology}, vol.~8, no.~2,
  p. 024005, feb 2023. [Online]. Available:
  \url{https://dx.doi.org/10.1088/2058-9565/acb6ae}
\BIBentrySTDinterwordspacing

\bibitem{mckay2022}
S.~Chen. (2022) Should we build quantum computers at all?
  \url{https://www.aps.org/publications/apsnews/202209/build-quantum.cfm}
  (Retrieved on: 27/02/2023).

\bibitem{campbell2023}
C.~Campbell. (2023) Quantum computers could solve countless problems—and
  create a lot of new ones.
  \url{https://time.com/6249784/quantum-computing-revolution/} (Retrieved on:
  27/02/2023).

\bibitem{deJong2022}
\BIBentryALTinterwordspacing
E.~de~Jong, ``Own the unknown: An anticipatory approach to prepare society for
  the quantum age,'' \emph{Digital Society}, vol.~1, no.~2, p.~15, Aug 2022.
  [Online]. Available: \url{https://doi.org/10.1007/s44206-022-00020-4}
\BIBentrySTDinterwordspacing

\bibitem{coenen2017}
\BIBentryALTinterwordspacing
C.~Coenen and A.~Grunwald, ``Responsible research and innovation (rri) in
  quantum technology,'' \emph{Ethics and Information Technology}, vol.~19,
  no.~4, pp. 277--294, Dec 2017. [Online]. Available:
  \url{https://doi.org/10.1007/s10676-017-9432-6}
\BIBentrySTDinterwordspacing

\bibitem{TQI2021}
TQI. (2021) Quantum ethics | a call to action.
  \url{https://thequantuminsider.com/2021/02/01/quantum-ethics-a-call-to-action/}
  (Retrieved on: 27/02/2023).

\bibitem{vermaas2017}
\BIBentryALTinterwordspacing
P.~E. Vermaas, ``The societal impact of the emerging quantum technologies: a
  renewed urgency to make quantum theory understandable,'' \emph{Ethics and
  Information Technology}, vol.~19, no.~4, pp. 241--246, Dec 2017. [Online].
  Available: \url{https://doi.org/10.1007/s10676-017-9429-1}
\BIBentrySTDinterwordspacing

\bibitem{meyer2023}
\BIBentryALTinterwordspacing
J.~C. Meyer, G.~Passante, S.~J. Pollock, and B.~R. Wilcox, ``How media hype
  affects our physics teaching: A case study on quantum computing,'' 2023.
  [Online]. Available: \url{https://arxiv.org/abs/2301.10882}
\BIBentrySTDinterwordspacing

\bibitem{meyer2022}
\BIBentryALTinterwordspacing
J.~Meyer, N.~Finkelstein, and B.~Wilcox, ``Ethics education in the quantum
  information science classroom: Exploring attitudes, barriers, and
  opportunities,'' 2022. [Online]. Available:
  \url{https://arxiv.org/abs/2202.01849}
\BIBentrySTDinterwordspacing

\bibitem{roberson2021}
\BIBentryALTinterwordspacing
T.~Roberson, ``Talking about responsible quantum: Awareness is the absolute
  minimum... that we need to do,'' 2021. [Online]. Available:
  \url{https://arxiv.org/abs/2112.01378}
\BIBentrySTDinterwordspacing

\bibitem{coenen2022}
\BIBentryALTinterwordspacing
C.~Coenen, A.~Grinbaum, A.~Grunwald, C.~Milburn, and P.~Vermaas, ``Quantum
  technologies and society: Towards a different spin,'' \emph{NanoEthics},
  vol.~16, no.~1, pp. 1--6, Apr 2022. [Online]. Available:
  \url{https://doi.org/10.1007/s11569-021-00409-4}
\BIBentrySTDinterwordspacing

\bibitem{memorandum2022}
S.~D. Young. (2022) Migrating to post-quantum cryptography.
  \url{https://www.whitehouse.gov/wp-content/uploads/2022/11/M-23-02-M-Memo-on-Migrating-to-Post-Quantum-Cryptography.pdf}
  (Retrieved on: 27/02/2023).

\bibitem{sandbox500}
M.~Swayne. (2023) Sandbox {AQ} announces \$500 million fundraise.
  \url{https://thequantuminsider.com/2023/02/15/sandbox-aq-announces-500-million-fundraise/}
  (Retrieved on: 27/02/2023).

\bibitem{qirg}
QIRG. (2020) Quantum internet research group (qirg).
  \url{https://datatracker.ietf.org/doc/charter-irtf-qirg/} (Retrieved on:
  27/02/2023).

\bibitem{etsiqkd}
ETSI. (2019) Quantum key distribution (qkd).
  \url{https://www.etsi.org/technologies/quantum-key-distribution} (Retrieved
  on: 27/02/2023).

\bibitem{canada}
``Overview of canada’s national quantum strategy,''
  \url{https://ised-isde.canada.ca/site/national-quantum-strategy/en},
  (Accessed on 06/30/2023).

\end{thebibliography}
\end{document}